\documentclass[letterpaper,number,1p]{elsarticle}

\usepackage[T1]{fontenc}
\usepackage[latin9]{inputenc}
\usepackage{amsmath}
\usepackage{amssymb}
\usepackage{esint}
\usepackage{microtype}
\usepackage{subfig}
\usepackage{booktabs}
\usepackage{amsfonts}
\usepackage{todonotes}
\usepackage{enumitem}
\usepackage{comment}
\usepackage{mathtools}
\usepackage{lineno}
\begin{document}
\begin{frontmatter}
\title{\Large Quasi-static Incremental Behavior of Granular Materials:
              Elastic--Plastic Coupling and Micro-scale Dissipation}
\author[up]{Matthew~R.~Kuhn\corref{cor1}}
  \ead{kuhn@up.edu}
\author[insa]{Ali~Daouadji}
  \ead{ali.daouadji@insa-lyon.fr}
\cortext[cor1]{Corresponding to:
               Donald P. Shiley School of Engineering,
               University of Portland,
               5000 N. Willamette Blvd.,
               Portland, OR, 97203, USA.
               Email: \texttt{kuhn@up.edu}}
\address[up]{Br. Godfrey Vassallo Prof. of Engrg.,
             Donald P. Shiley School of Engrg., Univ. of Portland,
             5000 N. Willamette Blvd., Portland, OR, USA 97231}
\address[insa]{University of Lyon, INSA-Lyon, GEOMAS, F-69621, France}
\begin{abstract}
The paper addresses a common assumption of elastoplastic modeling:
that the recoverable, elastic strain increment
is unaffected by alterations of the elastic moduli that
accompany loading.
This assumption is found to be false for a granular material,
and discrete element (DEM) simulations demonstrate
that granular materials are coupled materials at both
micro- and macro-scales.
Elasto-plastic coupling at the macro-scale
is placed in the context of thermomechanics
framework of Tomasz Hueckel and Hans Ziegler,
in which the elastic moduli are altered by irreversible processes
during loading.
This complex behavior is
explored for multi-directional loading probes that follow an
initial monotonic loading.
An advanced DEM model is used in the study, with
non-convex non-spherical particles and two different contact models:
a conventional linear-frictional model and an exact implementation of the
Hertz-like Cattaneo--Mindlin model.
Orthotropic true-triaxial probes were used in the study (i.e., no
direct shear strain), with tiny strain increments of $2\times 10^{-6}$.
At the micro-scale,
contact movements were monitored during small increments
of loading and load-reversal,
and results show that these movements are not reversed by a reversal of
strain direction, and
some contacts that were sliding during a loading increment
continue to slide during reversal.
The probes show that the
coupled part of a strain increment, the difference between the
recoverable (elastic) increment and its
reversible part,
must be considered when partitioning strain increments
into elastic and plastic parts.
Small increments of irreversible (and plastic) strain
and contact slipping and frictional dissipation
occur for all directions of
loading, and an elastic domain, if it exists
at all, is smaller than the strain increment used in the simulations.
\end{abstract}
\end{frontmatter}
%
%
\section{\large Introduction}
%
We use discrete element (DEM) simulations to address
the micro-scale nature of elastic and plastic deformation
in granular materials,
and to determine possible
coupling of the plastic and elastic responses
and its micro-scale origins.
The term ``elastic--plastic coupling'' is used
to describe a dependence of the elastic moduli on
alterations in a material's internal
fabric that are brought about by
irreversible processes that accompany plastic deformation
\cite{Hueckel:1976a,Ziegler:1983a,Gajo:2004a,Lashkari:2014a}.
Although elastic-plastic coupling is manifested in granular
materials as a degradation (and,in some cases, an augmentation)
of the elastic moduli,
the extent and nature of coupling are open issues,
ones that we explore in the paper.
%
\par
The remainder of this Introduction briefly presents the concepts of thermomechanics
that permit a distinction between elastic and reversible deformation
increments in
coupled materials.
The Introduction also explains the role of DEM simulations and the manner
in which they are used in the paper,
and it briefly describes the two models of contact friction that are used
in the simulations.
Section~\ref{sec:probes} focuses on tiny
increments of loading and characterizes
the micro-scale processes that occur during
these increments.
The section also presents methods for measuring
the small coupling strain increments that are the difference
between the elastic and reversible increments.
Similar methods are applied in Section~\ref{sec:multi}
to multi-directional probes that determine the constituent
parts of the incremental
strain~--- elastic, reversible, plastic, irreversible,
and coupled~--- that result
from small loading increments.
We find evidence of such elastic--plastic coupling
and of the contact-scale
dissipation that can enable this coupling,
particularly in certain directions of loading.
The results are verified for the two
models of contact interactions between particles.
%
\subsection{Thermomechanics of coupled materials}\label{sec:thermo}
In this section, we define the elastic and reversible constituents
of strain
increments, their plastic and irreversible counterparts,
and a coupled part that differentiates them.
These definitions are presented in the context
of continuum thermomechanics \cite{Ziegler:1983a,NematNasser:2013a},
of which the first law requires
that increments in the internal energy density, $dU$,
result
solely from mechanical stress-work, $\sigma_{ij}d\varepsilon_{ij}$,
and the external supply of heat, $dq$, per unit of volume,
\begin{equation}\label{eq:dU}
 dU = \sigma_{ij} d\varepsilon_{ij} + dq
\end{equation}
where $d\varepsilon_{ij}$ is a small-strain measure
of incremental deformation;
$\sigma_{ij}$ is the corresponding conjugate stress;
and density $U=U(\varepsilon_{ij}^{\text{(e)}},S)$ is taken
as a function of the current elastic strain
$\varepsilon_{ij}^{\text{(e)}}$ and of the entropy density $S$.
In this equation, summation applies to the $i$ and $j$ subscripts.
The cumulative elastic strain can, in principle, be measured as the
strain that is recovered upon fully unloading the material
to an unstressed reference configuration \cite{Hueckel:1976a}.
Incremental changes in the entropy are separated into reversible
and irreversible parts, $dS=dS^{\text{(r)}}+dS^{\text{(i)}}$,
of which the reversible part, $dS^{\text{(r)}}=dq/\theta$,
is the quotient of heat input and temperature $\theta$.
The second law holds that
\begin{equation}\label{eq:dS}
\theta dS \ge dq
\quad\text{or}\quad
\theta ds^{\text{(i)}} =
\theta\left(
dS - dS^{\text{(r)}}
\right) \ge 0
\end{equation}
\par
The Helmholtz free energy density $F$ is defined by the transformation
$F=U-\theta S$.
Henceforth, we consider only isothermal conditions ($d\theta=0$),
such that
\begin{equation}\label{eq:dF2}
dF = dU - \theta dS
\end{equation}
and combined with Eqs.(\ref{eq:dU}) and~(\ref{eq:dS}),
\begin{equation}\label{eq:thetaS}
  \sigma_{ij}d\varepsilon_{ij}-dF = \theta dS^{\text{(i)}} \ge 0
\end{equation}
which is a measure of dissipation for isothermal conditions.
The free energy is assumed a function of the elastic strain
$\varepsilon^{\text{(e)}}$ and a set of $K$ internal variables
$\alpha_{k}, k=1,2,\ldots,K$, that
are necessary to fully depict the current state
(see \cite{Coleman:1967a,Kestin:1970a,Mandel:1973a,%
Dafalias:1977b,Dafalias:1977c,Ziegler:1983a,Collins:1997a}):
%
$F = F(\varepsilon^{\text{(e)}},
\alpha_{1},\alpha_{2},\ldots,\alpha_{K})$,
%
noting that an explicit dependence upon temperature
is absent for the assumed isothermal conditions.
Changes in $F$ are produced by increments
$d\varepsilon_{ij}^{\text{(e)}}$ and $d\alpha_{k}$,
\begin{equation}\label{eq:dF1}
  dF = \frac{\partial F}{\partial\varepsilon_{ij}^{\text{(e)}}}
       d\varepsilon_{ij}^{\text{(e)}} +
       \frac{\partial F}{\alpha_{k}}d\alpha_{k}
\end{equation}
with summation over the $i$, $j$, and $k$ components.
The accumulated
total strain $\varepsilon_{ij}$ is the sum of elastic
and plastic parts,
$\varepsilon_{ij}=
\varepsilon_{ij}^{\text{(e)}} + \varepsilon_{ij}^{\text{(p)}}$,
the latter, in principle, being the unrecovered strain
of a deformed material after it is unloaded
to the unstressed reference condition.
Combining Eqs.~(\ref{eq:thetaS}) and~(\ref{eq:dF1}),
\begin{equation}
\left(\sigma_{ij}
      - \frac{\partial F}{\partial\varepsilon_{ij}^{\text{(e)}}}\right)
d\varepsilon_{ij}^{\text{(e)}}
+ \sigma_{ij}d\varepsilon_{ij}^{\text{(p)}}
- \frac{\partial F}{\alpha_{k}}d\alpha_{k} =
\theta dS^{\text{(i)}} \ge 0
\end{equation}
as in \cite{Coleman:1967a,Mandel:1973a}.
This inequality applies to all increments of
$d\varepsilon_{ij}^{\text{(e)}}$
and $d\alpha_{k}$, which implies that
\begin{align}\label{eq:sigmaF}
&\sigma_{ij} = \frac{\partial F}{\partial\varepsilon_{ij}^{\text{(e)}}}\\
&\sigma_{ij}d\varepsilon_{ij}^{\text{(p)}}
- \frac{\partial F}{\alpha_{k}}d\alpha_{k} \ge 0
\end{align}
the latter being the dissipation inequality.
A stress increment $d\sigma_{ij}$ is the sum
of parts attributed to changes in the elastic strain components
and to changes in the internal variables,
\begin{equation}\label{eq:dsigma}
d\sigma_{ij} =
\frac{\partial^{2}F}{\partial\varepsilon_{ij}^{\text{(e)}}
                     \partial\varepsilon_{kl}^{\text{(e)}}}
d\varepsilon_{kl}^{\text{(e)}} +
\frac{\partial^{2}F}{\partial\varepsilon_{ij}^{\text{(e)}}\partial\alpha_{k}}
d\alpha_{k}
\end{equation}
showing that alterations of the internal variables,
which typically accompany inelastic loading,
can modify the elastic moduli.
Hueckel \cite{Hueckel:1976a} classifies such
modifications as elastic hardening
or elastic softening, depending on whether the
the second term in Eq.~(\ref{eq:dsigma}) is aligned or
counter-aligned with the first term during a loading increment.
%
%
\par
Equation~(\ref{eq:sigmaF}) is inverted by using the Gibbs
free energy,
$G(\sigma_{ij},\alpha_{1},\alpha_{2},\ldots,\alpha_{K})$,
defined with the transformation
$G=-F+\sigma_{ij}\varepsilon_{ij}^{\text{(e)}}$.
By combining Eqs.~(\ref{eq:dU})--(\ref{eq:dF2}), and noting that
Eq.~(\ref{eq:thetaS}) applies to all increments $d\sigma_{ij}$,
we have
%
%
%
\begin{equation}\label{eq:varepsilon}
\varepsilon_{ij}^{\text{(e)}} =
\frac{\partial G}{\partial\sigma_{ij}}
\end{equation}
With this relation, an increment of
strain, $d\varepsilon_{ij}$, which is
the sum of elastic and plastic increments,
is given by
\begin{align}\label{eq:depsep}
d\varepsilon_{ij} =
d\varepsilon_{ij}^{\text{(e)}}
  + d\varepsilon_{ij}^{\text{(p)}} &=
  \frac{\partial^{2}G}{\partial\sigma_{ij}\partial\sigma_{kl}}
    d\sigma_{kl}
  + \frac{\partial^{2}G}{\partial\sigma_{ij}\partial\alpha_{k}}
    d\alpha_{k}
  + d\varepsilon_{ij}^{\text{(p)}}\\
  \label{eq:rcp}
  &=
  d\varepsilon_{ij}^{\text{(r)}}
  + d\varepsilon_{ij}^{\text{(c)}}
  + d\varepsilon_{ij}^{\text{(p)}}
\end{align}
In this separation of the total strain increment,
proposed by Hueckel, Maier, and Nova \cite{Hueckel:1977a,Hueckel:1979a},
the first constituent on the right is the reversible part
of the strain
increment, $d\varepsilon_{ij}^{\text{(r)}}$,
so named because it would be reversed
by a reversal of the stress increment,
$d\sigma_{ij}\leftrightarrow -d\sigma_{ij}$ in the absence
of any change in the internal condition
of the material, as expressed with the $\alpha_{k}$ variables.
The coupled constituent
$d\varepsilon_{ij}^{\text{(c)}}$ is the portion
of the elastic increment attributed to
alterations of the elastic moduli that accompany
irreversible processes, processes
that are brought forth by changes
in the internal variables:
a coupling of the elastic moduli to the plastic deformation.
\par
Changes of the internal variables are hidden during
standard laboratory tests of granular specimens,
as they are the consequence of micro-scale phenomena
such as crack growth within the grain fragments,
asperity fracture at the inter-granular contacts,
slip and micro-slip softening of the contact response,
and the general rearrangement of particles
that accompanies contact slip and micro-slip.
Although hidden in laboratory tests,
these micro-scale processes are exposed
(and can be tracked) in particle-scale numerical
simulations.
Notably, processes that would otherwise
change the internal variables $\alpha_{k}$ can be
momentarily frozen in simulations,
producing a constrained equilibrium state
that is nearby a preceding unconstrained state
\cite{Kestin:1970a,Rice:1971a} and
allowing separate measurements of the small incremental parts
$d\varepsilon_{ij}^{\text{(r)}}$,
$d\varepsilon_{ij}^{\text{(c)}}$, and
$d\varepsilon_{ij}^{\text{(p)}}$.
Our particular simulations were of durable (non-breaking)
particles, which could undergo slip and micro-slip
at their contacts and be rearranged as a result.
These irreversible processes were momentarily frozen
in a series of deformation probes
(Section~\ref{sec:probes}).
\par
\begin{figure}
  \centering
  \includegraphics{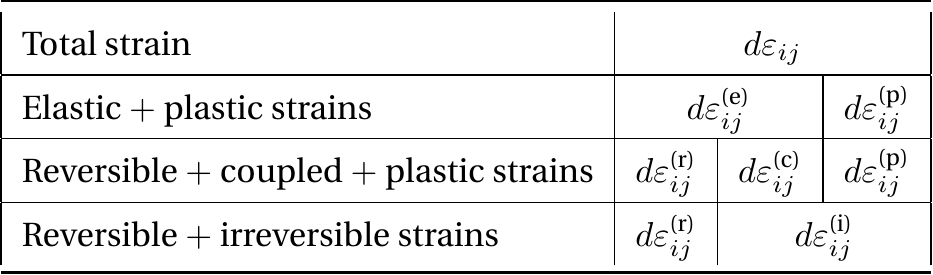}
  \caption{Decomposition of strain increments, from
           Collins and Houlsby \cite{Collins:1997a}.
           \label{fig:StrainDecomp}}
\end{figure}

Equation~(\ref{eq:rcp}) is regrouped as
\begin{equation}\label{eq:parts}
  d\varepsilon_{ij}^{\text{(e)}} =
  d\varepsilon_{ij}^{\text{(r)}} +
  d\varepsilon_{ij}^{\text{(c)}}, \quad
  d\varepsilon_{ij}^{\text{(i)}} =
  d\varepsilon_{ij}^{\text{(c)}} +
  d\varepsilon_{ij}^{\text{(p)}}, \quad\text{and}\quad
  d\varepsilon_{ij} =
  d\varepsilon_{ij}^{\text{(r)}} +
  d\varepsilon_{ij}^{\text{(i)}}
\end{equation}
in which the elastic--plastic coupling part
$d\varepsilon_{ij}^{\text{(c)}}$ is a part of the elastic
increment, and the irreversible increment
$d\varepsilon_{ij}^{\text{(i)}}$ is the part
of the total increment $d\varepsilon_{ij}$
that, in the absence of changes $d\alpha_{k}$, is not
reversed by a reversal of the stress increment
$d\sigma_{ij}$
\cite{Hueckel:1977a,Hueckel:1979a}.
This decomposition of strains is summarized in
Fig.~\ref{fig:StrainDecomp}, as
adapted from \cite{Collins:1997a}.
\subsection{DEM modeling in relation to a thermomechanics framework}%
\label{sec:DEMadvantage}
Discrete particle (DEM) simulations were used to measure
strain increments for different directions of loading
and to decompose an increment of total strain into the
various parts from which it is derived:  the
reversible, irreversible, elastic, plastic, and coupled parts.
We also used simulations
to gain an understanding of the
underlying micro-scale mechanisms that contribute
to bulk deformation and of contact-scale frictional dissipation.
Because micro-scale information and more
general boundary conditions are accessible with
DEM simulations, they can expose granular behavior in ways
that are not easily achieved with laboratory testing.
Specifically,
four advantages of DEM simulations were
exploited in our simulations:
\begin{enumerate}[listparindent=2em]
\item
After a granular assembly has been created and loaded to
a given state, this state and all of its micro-data
(particle positions, contact forces, J\"{a}ger-equivalent
contact history, etc.)
can be saved.
This saved state~--- which Alonso-Marroqu\'{i}n \cite{AlonsoMarroquin:2005b}
calls a numeric ``clone''~--- constitutes a consistent
reference state for subsequent loading probes that
precisely captures (to within machine precision)
the essential micro-scale condition of the system.
\item
In performing loading probes,
DEM codes permit control of any
combination of six components of the
stress and strain tensors or any six linearly independent
combinations of stress
and strain~--- loading
conditions that would require multiple sample preparations
and even different testing apparatus in a physical laboratory
\cite{Kuhn:2002b}.
Such multi-directional probes are used to fully characterize
elastic--plastic coupling
(Section~\ref{sec:multi}).
\item
One can extract the bulk
elastic energy and the bulk frictional
dissipation from their micro-scale origins within a material
(Section~\ref{ref:probe1}).
The ``stored'' energy
is computed at each inter-particle contact,
and the total contact energy of all contacts is
a direct measure of the Helmholtz energy $F$
and any incremental change $dF$.
\par
Micro-scale contact frictional dissipation can also be obtained
from these probes.
Dissipation in granular materials can result from
frictional slip, frictional micro-slip,
plastic deformation of the particles,
particle fracture, and viscous drag of the particles as they
move through (or relative to) a pore fluid.
Our focus is on the slow,
quasi-static
loading of uncemented dry materials
at low confining pressure
(the mean stress,
defined as the negative of $1/3$ of the first stress invariant)
and with durable particles,
for which frictional contact slip and micro-slip are
the dominant dissipation mechanisms.
As such,
viscous damping
is kept as small as is necessary to maintain computational
stability,
so that dissipation results solely from
frictional slip and micro-slip at the contacts
(Section~\ref{ref:probe1} and \ref{sec:DEM}).
\item
By momentarily suspending contact slip and micro-slip
during a loading probe,
DEM simulations can be used for distinguishing
the reversible and irreversible parts of a total
strain increment.
When compared with a conventional, unconstrained probe,
one can decompose the total increment $d\varepsilon_{ij}$
into the source parts of Eqs.~(\ref{eq:rcp})
and~(\ref{eq:parts})
(see the Types~2 and~3 probes
of Section~\ref{sec:probes}).
\end{enumerate}
%
Although simulations offer these advantages,
the results depend, however, upon the model
that is ascribed to the individual contacts, an issue that is
addressed in our study.
%
\subsection{Two contact models:
            linear-frictional and Cattaneo--Mindlin}\label{sec:models}
%
The 
deformation of a granular material with durable grains
depends upon
the material's contact-scale behavior,
and two contact models are used in this study:
a standard linear-frictional contact model and a
Hertz--Mindlin model,
which we will call the \emph{Cattaneo--Mindlin} contact.
The purpose is to ascertain
the bulk incremental behavior
and its relation to the particular contact model.
Both models are elastic with respect to movements
that are normal to the particles' surfaces:
the first model has a linear force-displacement relation
for normal movements;
the second has a non-linear Hertz-type relation.
Moreover,
because the contact loads and indentations
are small, the contact zones are finite but small, and the behavior
of a single contact is unaffected by the loading of more distant
contacts on the same particle
(see \cite{Gonzalez:2012a} for the case of large indentations).
\par
For tangential movements,
the standard linear-frictional model uses a linear force-dis\-place\-ment
relation, 
with elastic ``spring'' stiffnesses
$k^{\text{n}}$ and $k^{\text{t}}$
in the normal and tangential directions.
The behavior is entirely elastic when
$|\mathbf{f}^{\text{t}}|<\mu f^{\text{n}}$,
where $f^{\text{n}}$ and $\mathbf{f}^{\text{t}}$
are the normal and tangential forces,
and $\mu$ is the contact friction coefficient.
Contact behavior can become partially or wholly plastic
(i.e., with slip)
when the friction limit is
reached, $|\mathbf{f}^{\text{t}}|=\mu f^{\text{n}}$,
and the frictional loss (energy dissipation) can
be computed for a sliding contact.
Details of this model's incremental behavior
are given in~\ref{app:contacts}. 
\par
A second set of simulations used
a more realistic contact model
having a Hertz-like and elastic normal contact stiffness
and an exact representation of the corresponding inelastic
tangential response.
Hertz contact is recognized as an improved
representation of elasto-frictional particle interaction,
but
because of its coding complexities and computational demands,
a faithful implementation of Hertz contact, which should include
micro-slip, is rarely fulfilled in DEM simulations.
Two isotropic elastic spheres, 
when pressed together,
touch within a circular contact patch
between the particles' two 
surfaces.
The corresponding inelastic tangential force-displacement
behavior is modeled with Cattaneo--Mindlin theory
\cite{Mindlin:1953a}.
When a small tangential force is applied between the two spheres,
frictional slip does not
occur uniformly across the full
contact patch, instead, the slip is confined to an outer annular
ring with a width that depends on the applied tangential force
and on the force (or displacement) history.
Such annular ``micro-slip'' produces frictional dissipation
within the contact patch,
which can give way to full sliding
of the entire contact patch when the tangential force
$|\mathbf{f}^{\text{t}}|$ reaches the limit $\mu f^{\text{n}}$
(i.e., 
the outer slip cone $|\mathbf{f}^{\text{t}}|=\mu f^{\text{n}}$
in Fig.~\ref{fig:cones}c).
Because micro-slip and dissipation
can occur inside this slip cone,
the contact can exhibit plastic and incrementally nonlinear
behaviors when the tangential
force \emph{is less than} the friction limit
(when $|\mathbf{f}^{\text{t}}|<\mu f^{\text{n}}$).
This response is akin to elasto-plastic strain hardening
at the contact scale.
The resulting incremental contact behavior can be elastic,
micro-slip, or full slip, depending upon the
direction of loading.
More commonly, approximate implementations of the Hertz contact
simply
use a tangential stiffness that is proportional
to the normal force and not upon the directions of
the normal and tangential movements \cite{Cundall:1988a,Lin:1997a}.
Such implementations do not permit frictional dissipation in the form
of micro-slip and can result in an unfortunate and physically
inadmissible infusion of energy during close cycles of
contact movement \cite{Elata:1996a}.
Details of the exact implementation of
the Cattaneo--Mindlin model used in this study are
described in~\ref{app:contacts} 
and the references therein.
\par
The two contact models can be summarized
in the context of a thermomechanic framework.
Elastic energy is stored within the contacts
and frictional dissipation is produced
by slip and micro-slip, and both incremental
consequences of the stress-work $\sigma_{ij}d\varepsilon_{ij}$
can be audited throughout a DEM simulation.
Besides the shapes of the particles and their
arrangements (i.e., the material's internal geometric fabric),
the internal condition of a material with simple linear-frictional
contacts is additionally described by the current normal and
tangential forces at each contact.
A material with Cattaneo--Mindlin contacts
requires this information
and more, including the
full J\"{a}ger equivalent-loading information of each contact
(\ref{app:contacts}).
All of this information can be stored in DEM files,
so that each probe in a series of probes can
retrieve the information, and that each probe is initiated from
the same internal condition.
%
\section{\large Monotonic triaxial compression}%
\label{sec:monotonic}
We used a series of slow, quasi-static
DEM simulations to study the incremental response of
a granular assembly.
The assembly was a cubical box filled with 10,648
smooth non-convex sphere-cluster particles
and contained within
periodic boundaries.
The particles' shape, sizes, and arrangement were
calibrated to simulate the behavior of a
fine-grain poorly-graded medium-dense sand
(Nevada Sand \cite{Arulmoli1992a,Kuhn:2014c};
model details are described in \ref{sec:DEM}). 
The assembly was large enough to
capture the average material behavior but sufficiently small to
prevent meso-scale localization, such as shear bands.
The initial particle arrangement was isotropic with a confining
pressure,
$-\tfrac{1}{3}(\sigma_{11}+\sigma_{22}+\sigma_{33})$, of 100~kPa.
\par
Although it is inconsistent with the quasi-static hypothesis,
DEM results can depend upon
the loading rate~--- an unfortunate dependence which is infrequently
discussed in a candid manner in the DEM literature.
To mitigate such rate dependence,
a slow loading rate and other measures
were taken, assuring that the simulations were nearly
quasi-static.
Because of their importance in the current study,
computational essentials are detailed in \ref{sec:DEM}, 
which also
describes quantitative indicators verifying the quasi-static nature
of the simulations.
\par
\begin{figure}
  \centering
  \includegraphics{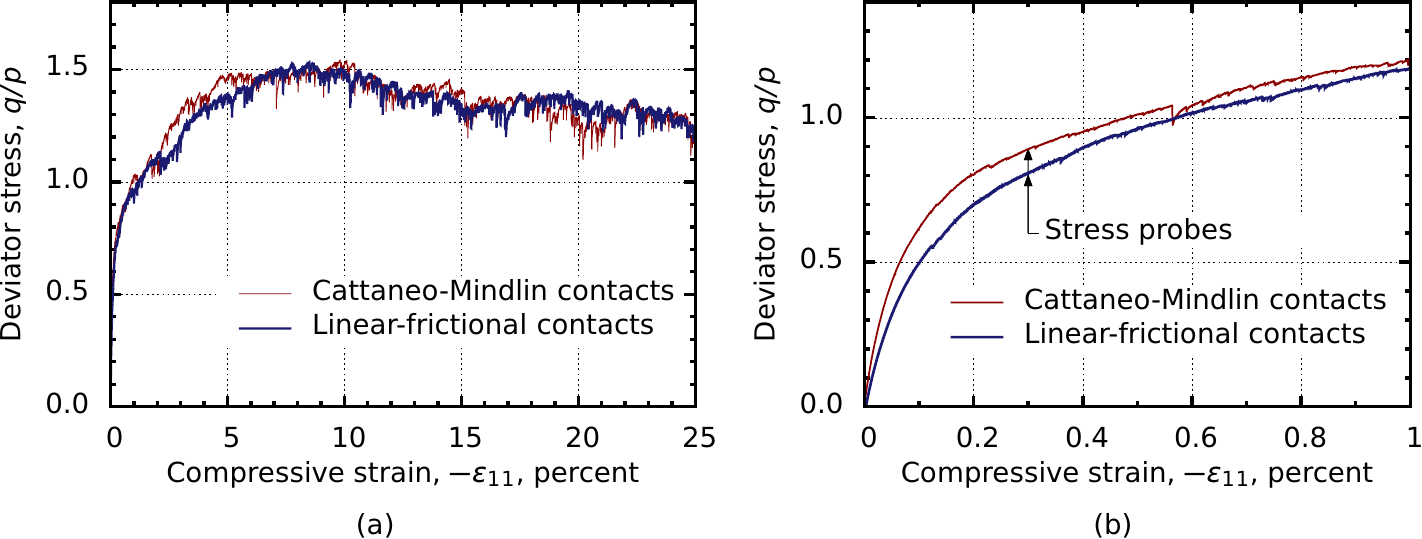}
  \caption{Stress and strain during DEM simulation of
           triaxial compression
           at constant mean stress with
           10,648 non-convex sphere-cluster particles
           ($p_{\text{o}}=100$~kPa).
           Stress probes in Sections~\ref{sec:probes}
           and~\ref{sec:multi} were initiated from a strain
           of 0.3\%.
           \label{fig:StressStrain}}
\end{figure}

Figure~\ref{fig:StressStrain} shows the results of
drained constant-$p$ triaxial compression,
in which the $x_{1}$ width
of the assembly was reduced at constant rate $\dot{\varepsilon}_{11}$
while adjusting the lateral widths to maintain a constant mean stress
$p$ of 100~kPa.
Henceforth, this type of axisymmetric
triaxial loading is referred to as constant-$p$ loading.
Results are shown for the two contact models:
linear--frictional and Cattaneo--Mindlin.
The initial, small-strain modulus differed
for the two simulations 
(initial moduli of elasticity $E$
of 83~GPa and 111~GPa
for the linear-frictional
and Cattaneo--Mindlin models respectively).
Better agreement of these small-strain moduli could be obtained,
of course, by further tuning the contact stiffnesses,
but the results are considered satisfactory
for the current study.
Results with the two contact models
were quite similar, however,
when the strains were larger than 0.5\%.
This similarity
indicates that the large-strain monotonic behavior is
less sensitive to the details
of the contact model, provided that different
models incorporate the same
friction coefficient $\mu$.
%
%
%
\section{\large Stress probes, dissipation, and coupled response}\label{sec:probes}
\begin{figure}
  \centering
  \includegraphics{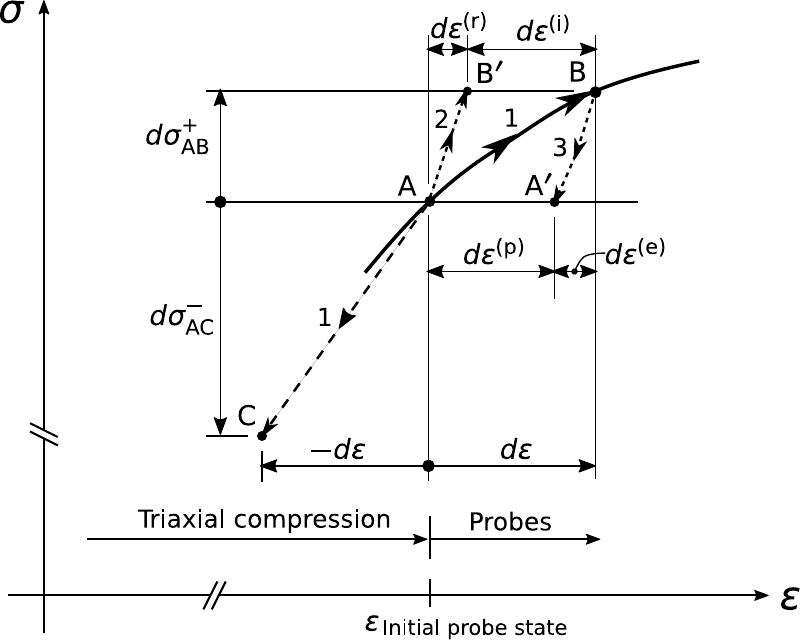}
  \caption{Three pairs of loading probes.
           Numerals correspond to the Type~1,
           Type~2, and Type~3 probes that
           are summarized in Table~\ref{table:probes}.%
           \label{fig:probes}}
\end{figure}

DEM simulations were used to explore micro-scale
conditions at the contacts and to quantify the elastic,
plastic, reversible, irreversible, and coupled
parts of strain increments.
To these ends, we conducted
three types of stress and strain probes:
(1) loading-retrogression pairs,
(2) loading and reversible loading pairs, and
(3) loading-unloading pairs.
The three types of probes are illustrated in Fig.~\ref{fig:probes},
summarized in Table~\ref{table:probes},
and described in separate sections below.
The loading-retrogression pairs (Type~1) involved
a loading probe and a strain-reversal (retrogression) probe and
were used to compare the micro-scale changes in contact behavior
during bulk strains applied in opposite directions.
The loading and reversible loading pairs (Type~2) were
used to determine the reversible and irreversible
parts of a strain increment; whereas,
loading-unloading pairs (Type~3) produced
a small closed path in stress-space and were used to determine
the elastic and plastic parts of a strain increment.
Simulations with the two contact models,
Cattaneo--Mindlin and linear-frictional,
were also compared with these probes.
\begin{table}
  \caption{Summary of simulation probes.
           \label{table:probes}}
  \centering\small
  \begin{tabular}{clcll}
  \toprule
  Type & Probe pairs & Fig.~\ref{fig:probes} & Description & \\
  \midrule
  1    & Loading \& retrogression & A--B & reference loading increment &
         $d\boldsymbol{\varepsilon}\rightarrow
          d\boldsymbol{\sigma}^{+}_{\text{AB}}$ \\ 
       & & A--C & retrogression (strain-reversal) &
         $-d\boldsymbol{\varepsilon}\rightarrow
          d\boldsymbol{\sigma}^{-}_{\text{AC}}$ \\
          &&& increment & \\
  \midrule
  2 & Loading \& reversible loading& A--B & reference loading increment &
      $d\boldsymbol{\varepsilon}\rightarrow
          d\boldsymbol{\sigma}^{+}_{\text{AB}}$ \\
   &  & A--B$^\prime$ & reversible increment &
      $d\boldsymbol{\sigma}^{+}_{AB}\xrightarrow{\mu\rightarrow\infty}
       d\boldsymbol{\varepsilon}^{\text{(r)}}$ \\
   & & & & $d\boldsymbol{\varepsilon}^{\text{(i)}} =
            d\boldsymbol{\varepsilon} -
            d\boldsymbol{\varepsilon}^{\text{(r)}}$ \\
  \midrule
  3 & Loading \& unloading & A--B & reference loading increment &
      $d\boldsymbol{\varepsilon}\rightarrow
          d\boldsymbol{\sigma}^{+}_{\text{AB}}$ \\
   &  & B--A$^\prime$ & unloading increment &
      $-d\boldsymbol{\sigma}^{+}_{AB}\rightarrow
       -d\boldsymbol{\varepsilon}^{\text{(e)}}$ \\
   &  & & &
      $d\boldsymbol{\varepsilon}^{\text{(p)}} =
       d\boldsymbol{\varepsilon} -
       d\boldsymbol{\varepsilon}^{\text{(e)}}$ \\
  \bottomrule
  \end{tabular}
\end{table}
\par
The initial, reference states for the probes were established
with drained monotonic
constant-$p$ triaxial compression,
in which the mean stress $p$ was maintained constant during
axial ($x_{1}$ direction) loading
(Fig.~\ref{fig:StressStrain}).
At several strains,
we saved the entire state of all particles and contacts,
so that the different deformation probes started from
the same reference state~--- the ``initial probe state''
(point~A in Fig.~\ref{fig:probes}).
For all three types of probe-pairs,
the \emph{loading} probe was simply
a continuation of the previous triaxial compression:
starting
from the initial probe state, the strain $\varepsilon_{11}$ was
reduced (compressed) by an additional
$2\times 10^{-6}$ (path~A--B in Fig.~\ref{fig:probes}),
in a continuation of constant-$p$ compression.
A similar strain increment of $2\times 10^{-6}$
was used with all three types of probes and in various
loading directions.
Halving or doubling the strain increment or strain rate had minimal
effect on the results
(\ref{app:contacts}).
%
\subsection{Loading and retrogression (Type~1)
            probes with Cattaneo--Mindlin contacts}\label{ref:probe1}
The loading phase of these probes was begun
from an initial probe state
(the increment~AB in Fig.~\ref{fig:probes}).
For the results of this section, the loading was of continued
constant-$p$
triaxial compression with a strain increment of
$d\varepsilon_{11}=-2\times 10^{-6}$;
whereas, in Section~\ref{sec:multi} the loading
phase was conducted in other directions.
We noted and stored all three components of the
strain during this loading probe~---
components $d\varepsilon_{11}$,
$d\varepsilon_{22}$, and $d\varepsilon_{33}$~---
and starting again from the same initial probe state (point~A),
we precisely reversed these strain components for the
\emph{retrogression} (strain-reversal) probe~AC,
with the intent of determining the alterations of
micro-scale contact movements that were produced
by strain increments in opposite directions.
\par
\begin{figure}
  \centering
  \includegraphics{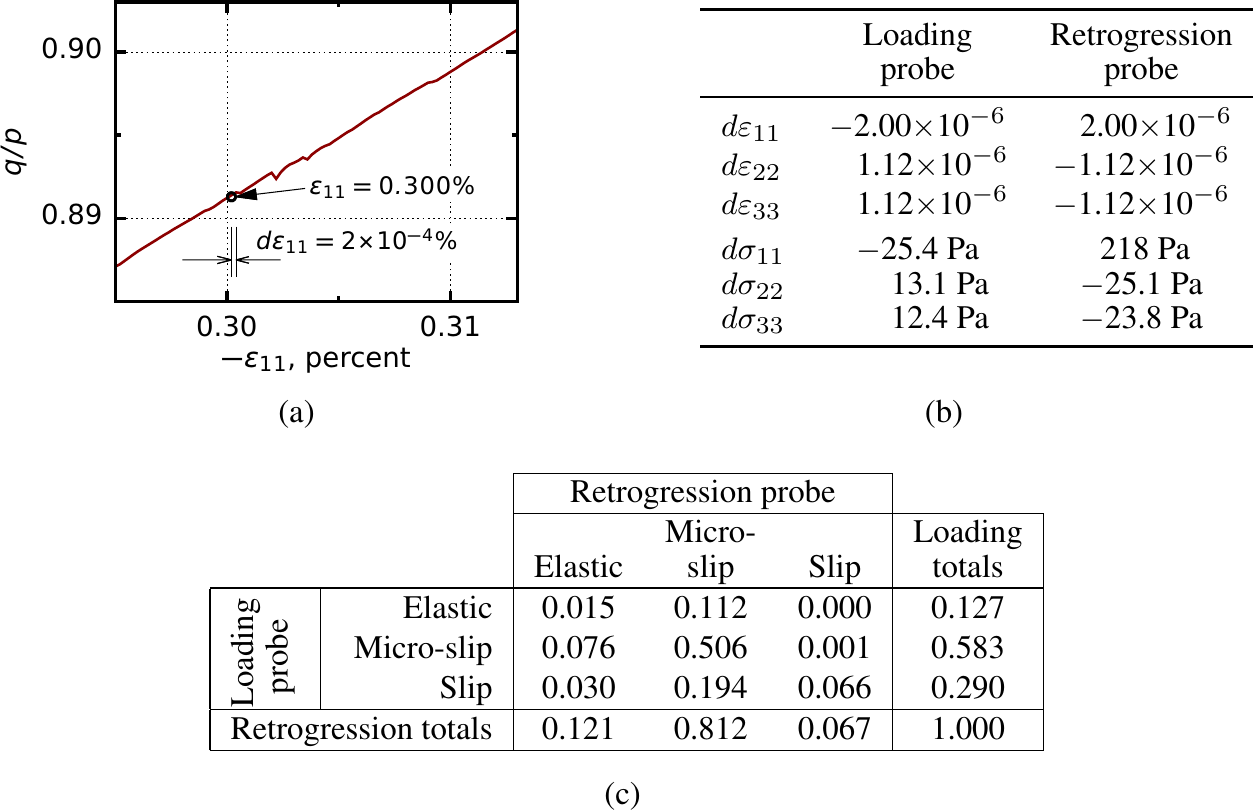}
  \caption{Results of loading and retrogression (strain-reversal)
           Type~1 probes, starting at strain
           $\varepsilon_{11}=-0.3\%$ with
           Cattaneo--Mindlin contacts.
           (a) stress-strain during constant-$p$ triaxial compression,
           with the small probe $d\varepsilon_{11}$
           (see Fig.~\ref{fig:StressStrain} for the full
           range of stress and strain);
           (b) strain and stress increments during loading (AB)
           and strain-reversal (AC) probes; and
           (c) fractions of contacts that
           transition among the three contact conditions
           (elastic, micro-slip, and slip) between
           the loading (AB) and strain-reversal (AC) probes.
           \label{fig:Probe1}}
\end{figure}

Figure~\ref{fig:Probe1} shows results for the
loading and retrogression probes that originated
at strain $\varepsilon_{11}=-0.3\%$ in the simulation
with Cattaneo--Mindlin (i.e., Hertz-type) contacts.
%
This initial probe state was during the strain-hardening phase,
when deviator stress was about 55\% of the peak strength
(Fig.~\ref{fig:Probe1}a).
As expected,
the stiffness was much smaller during loading than during
retrogression,
with moduli $d\sigma_{11}/d\varepsilon_{11}$ of 12.7~GPa and
109~GPa, respectively
(Fig.~\ref{fig:Probe1}b).
During the loading probe, the material dilated slightly, at rate
$dv/|d\varepsilon_{11}|=0.12$.
\par
By post-processing the micro-scale data, the condition of each
Cattaneo--Mindlin
contact~--- elastic,
micro-slip, or full slip~--- was
determined during loading and retrogression
(see Section~\ref{sec:models} and \ref{app:contacts}). 
Nine transitions are possible among
these three contact conditions, and
the table in Fig.~\ref{fig:Probe1}c
gives the fractions of contacts that
transitioned in each manner
between loading (rows) and retrogression (columns).
For example, 11.2\% of all contacts were elastic during
the loading increment but exhibited micro-slip during
the strain-reversal increment.
Eight of
the nine possible pairings occurred among the contacts.
Consider, for example, the 29.0\% of contacts that were
sliding (full slip) during loading.
Of these sliding contacts, 23\% ($=0.066/0.290$) were also
sliding during the strain-reversal (retrogression) probe,
67\% transitioned to
micro-slip, and only 10\% became fully
elastic.
Of the 12.7\% of contacts that were elastic during loading,
88\% transitioned to micro-slip
during the retrogression probe.
Of the contacts that were undergoing micro-slip, only 13\%
transitioned to an elastic condition during the strain-reversal probe,
and a small number had transitioned to full slip.
In other words, a reversal of strain does not
necessarily reverse (toggle)
the mode of a contact's movement:
some contacts that were sliding or in micro-slip continued
in these modes when the strain was reversed.
The only transition that was not observed was the unlikely
case of a contact that was elastic during loading but transitioned
to full slip during retrogression.
\par
Considering the small magnitude of the strain increment
($|d\varepsilon_{11}|=2$$\times$10$^{-6}$),
these results suggest that a zone of
purely reversible response
(or a cone, in the sense of Darve \cite{Darve:1990a})
does not exist during the strain-hardening phase.
Our results with linear-frictional contacts are given
in Section~\ref{sec:linfricprobe}, where a similar conclusion
is made.
This difficulty of distinguishing a finite zone of purely
reversible response
has also been noted in the simulations of
ductile powder compacts \cite{Harthong:2012a},
was anticipated by Hill \cite{Hill:1967a}
for multi-slip crystal plasticity, and is modeled
in the generalized plasticity of Mroz et al. \cite{Mroz:1978a}.
The results in this section are for a single pair
of loading and retrogression increments.
These results, by themselves, do not refute the possibility
of an elastic zone
(indeed, the generalized plasticity of
Mroz admits an elastic zone of zero measure at
the interface of adjacent plastic zones).
However, the multi-direction
probes of Section~\ref{sec:multi} show that
small plastic (and irreversible) increments
occur for all directions of loading.
\par
\begin{figure}
  \centering
  \includegraphics{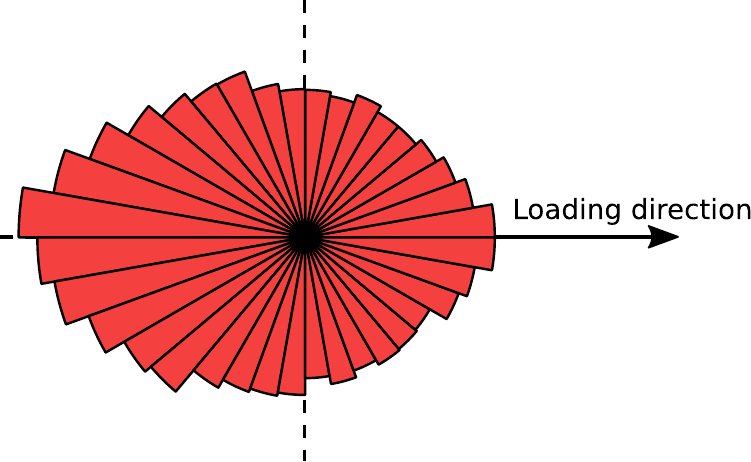}
  \caption{Comparison of the directions of tangential contact movements
           $d\xi$
           during loading and retrogression
           (strain-reversal) for Type~1 probes, initiated at
           at the reference strain
           $\varepsilon_{11}=-0.3\%$ with
           Cattaneo--Mindlin contacts.
           Each radial bar represents the number of contacts
           that, although moving toward the right during loading,
           were moving in the bar's direction during a reversal
           of the strain.
           \label{fig:Rose}}
\end{figure}

As a further investigation of contact behavior during these small
increments of loading and retrogression,
we compared the \emph{directions}
of the contact movements during the two probes that
began at strain $\varepsilon_{11}=-0.3\%$
(again, with simulations having
Cattaneo--Mindlin contacts).
Focusing on the tangential movements $d\boldsymbol{\xi}$ of pairs of particles
at their contacts, the radial bars in
Fig.~\ref{fig:Rose} show the directions in which
tangential movement occurred during
the strain-reversal (retrogression) probe
when compared with the
directions during the loading probe for the same contacts.
%
The arrow pointing to the right represents the movement direction
during loading
(the movement direction
of each contact during loading has been
rotated to this arrow's direction so that their
corresponding retrogression directions can be compared).
Bar sizes indicate the relative numbers of contacts
with a given movement direction.
The figure shows that during strain-reversal,
the directions of contact
movement are not simply
reversed from their original loading directions:
the movements are not strongly directed toward the left in
this figure.
In fact, there is only a small correlation
between the directions of contact movements
for the two opposite directions of bulk strain,
with only a small bias toward a reversal of the movements
when strain is reversed.
Quite simply,
micro-scale contact movements
are not reversed by a reversal of
the macro-scale deformation.
\par
To better understand the degree of irreversible behavior
during the retrogression
probe, we also audited energy changes, measuring
the increments in work density attributed
to elastic deformations at the contacts
and to dissipative, frictional micro-slip and slip at the
contacts.
For the loading probe, the increment of stress-work density
$\sigma_{ij}d\varepsilon_{ij}$
was a positive $0.82\,p\,d\varepsilon_{11}$
($p$ being the mean stress of 100kPa,
and $d\varepsilon_{11}$ being the positive increment
2$\times$10$^{-6}$);
and since deformation was precisely reversed
during the retrogression probe,
the corresponding increment was a negative
$-0.82\,p\,d\varepsilon_{11}$.
During the loading probe, frictional contact dissipation was
$0.79\,p\,d\varepsilon_{11}$, accounting
for about 96\% of the stress-work,
the remaining 4\% being a small change in the contacts'
elastic energy.
Although frictional contact
dissipation was less during the reversal probe
(about $0.13\,p\,d\varepsilon_{11}$), the
frictional dissipation
was still quite substantial, about 17\% of the stress-work,
while the elastic energy was reduced by
$-0.95\,p\,d\varepsilon_{11}$.
In short, the material exhibited
frictional dissipation in the form of contact
slip and micro-slip during both loading and
retrogression,
even with the very small strain increment of 2$\times$10$^{-6}$.
%
%
\subsection{Loading and reversible loading (Type~2)
            probes with Cattaneo--Mindlin contacts}\label{sec:Type2CM}
With Type~2 probes,
we determined the amount and direction of the reversible
and irreversible parts of a strain increment,
using a technique pioneered by Calvetti and his co-workers
\cite{Calvetti:2003a,Tamagnini:2005a}.
As we have seen, frictional contact
dissipation occurs both during loading and
during a reversal of the loading strain (retrogression).
Because such frictional dissipation indicates
that deformation is only partially
reversible during both loading and retrogression,
we can not separate the loading increments into
reversible and irreversible parts by simply
comparing pairs of loading-retrogression (Type~1) probes
or by using closed stress-paths (the latter are Type~3 probes,
which are considered below for another purpose).
To determine the reversible part of deformation,
we conducted a ``reversible probe'' (AB$^\prime$
in Fig.~\ref{fig:probes} and Table~\ref{table:probes})
by replicating the same \emph{stress increment} as during
loading (increment
$d\boldsymbol{\sigma}^{+}_{\text{AB}}$),
but by momentarily intervening (for the duration of the
strain increment) and preventing
micro-slip and slip
by assigning a large
friction coefficient to the contacts (i.e., $\mu=50$).
That is,
we first determined the stress increment
$d\boldsymbol{\sigma}^{+}_{\text{AB}}$ that resulted
from unadulterated loading with $\mu=0.55$
(the stresses reported in Fig.~\ref{fig:Probe1}b),
and then we produced this same stress increment
with $\mu=50$.
The latter strain increments
were free of slip and micro-slip
and were entirely the result of elastic contact
movements: these strain increments are the reversible
parts $d\boldsymbol{\varepsilon}^{\text{(r)}}$
in Eqs.~(\ref{eq:depsep})--(\ref{eq:rcp}),
and the irreversible parts $d\boldsymbol{\varepsilon}^{\text{(i)}}$
are the differences
$d\boldsymbol{\varepsilon}-d\boldsymbol{\varepsilon}^{\text{(r)}}$,
as in Eq.~(\ref{eq:parts}\textsubscript{3}).
\par
The results for a probe of constant-$p$ triaxial
compression are presented in Table~\ref{table:ElPlProbe},
showing the reversible and irreversible strain parts
of the loading probe.
%
\begin{table}
  \caption{Results of loading and reversible
           (Type~2) probes, starting at strain
           $\varepsilon_{11}=-0.3\%$ with
           Cattaneo--Mindlin contacts.
           Reversible strains were determined
           by artificially suppressing contact sliding.
           \label{table:ElPlProbe}}
  \centering\small
  \begin{tabular}{lcrrr}
    \toprule
    & Strain &
          \multicolumn{1}{c}{Reversible}
        & \multicolumn{1}{c}{Irreversible}
        & \multicolumn{1}{c}{Total} \\
    Probe & component &
          \multicolumn{1}{c}{$d\varepsilon^{\text{(r)}}$}
        & \multicolumn{1}{c}{$d\varepsilon^{\text{(i)}}$}
        & \multicolumn{1}{c}{$d\varepsilon$} \\
    \midrule
    Loading & $d\varepsilon_{11}$ &
             $-$2.94$\times$10$^{-7}$ & $-$17.06$\times$10$^{-7}$
             & $-$20.00$\times$10$^{-7}$\\
            & $dv$ &
            2.00$\times$10$^{-7}$ & 0.31$\times$10$^{-7}$
            & 2.31$\times$10$^{-7}$\\
    \bottomrule
  \end{tabular}
\end{table}
Both axial $d\varepsilon_{11}$ and volumetric $dv$
strains are reported.
The table shows
that over 85\% of the loading
strain $d\varepsilon_{11}$ was irreversible.
During loading, both reversible and irreversible parts of
volume change,
$dv^{\text{(r)}}$ and $dv^{\text{(i)}}$,
were dilative, with most of the incremental
dilation being reversible.
%
%
%
\subsection{Loading and unloading (Type~3)
            probes with Cattaneo--Mindlin contacts}\label{sec:type3a}
With the final variant of probes,
we used conventional pairs of loading and
unloading increments, producing a tiny closed path in stress-space
(Fig.~\ref{fig:probes} and Table~\ref{table:probes}).
This conventional technique is used in laboratory and numerical testing
to determine the elastic and plastic parts of a strain increment
(for example,
\cite{Bardet:1994a,Anandarajah:1995a,Royis:1998a,Kishino:2003a,AlonsoMarroquin:2005a}).
The loading probe AB is detailed in
Fig.~\ref{fig:Probe1}b for the initial probe state
at strain $\varepsilon_{11}=-0.3\%$, and this
loading increment advanced the compressive
strain $d\varepsilon_{11}$ by 2$\times$10$^{-6}$ while maintaining
constant mean stress, ending with the stress increment
$d\boldsymbol{\sigma}^{+}_{AB}$.
The unloading probe BA$^{\prime}$ was a reversal
$-d\boldsymbol{\sigma}^{+}_{AB}$
of the \emph{stress}
produced during forward loading, resulting in the elastic
strain increment $-d\varepsilon^{\text{(e)}}$
(that is, we reversed the stresses shown in the bottom
half of the loading column in Fig.~\ref{fig:Probe1}b).
This elastic increment includes a reversible part~--- the
increment
$(\partial^{2}G/\partial\sigma_{ij}\partial_{kl})d\sigma_{kl}$
in Eq.~(\ref{eq:depsep})~--- and
a coupled part that arises from changes
in the elastic moduli that accompany the plastic deformation.
The coupling part is associated with
the slip and micro-slip of contacts and small particle
rearrangements,
which occur during both loading and unloading.
With the pair of probes, we determined the elastic and plastic
strain increments, as given on the left of Eq.~(\ref{eq:depsep}).
\par
Table~\ref{table:epirc} summarizes the elastic and
plastic strain increments
for a probe of constant-$p$ compression
and also includes the reversible and irreversible
parts taken from Table~\ref{table:ElPlProbe}.
\begin{table}
  \centering\small
  \caption{Results of loading and unloading
           (Type~3) probes, starting at strain
           $\varepsilon_{11}=-0.3\%$ with
           Cattaneo--Mindlin contacts.
           Results of Type~2 probes are also shown.\label{table:epirc}}
  \begin{tabular}{lrr}
    \toprule
    & \multicolumn{2}{c}{Strain components}\\
    \cmidrule(l){2-3}
    Decomposition
      & \multicolumn{1}{c}{$d\varepsilon_{11}$}
      & \multicolumn{1}{c}{$dv$} \\
    \midrule
    Reversible, $d\varepsilon_{ij}^{\text{(r)}}$
                & $-$2.94$\times$10$^{-7}$
                & 2.00$\times$10$^{-7}$ \\
    Irreversible, $d\varepsilon_{ij}^{\text{(i)}}$
                & $-$17.06$\times$10$^{-7}$
                & 0.31$\times$10$^{-7}$\\
    \midrule
    Elastic, $d\varepsilon_{ij}^{\text{(e)}}$
                & $-$2.57$\times$10$^{-7}$
                &    2.37$\times$10$^{-7}$\\
    Plastic, $d\varepsilon_{ij}^{\text{(p)}}$
                & $-$17.43$\times$10$^{-7}$
                & $-$0.06$\times$10$^{-7}$\\
    \midrule
    Coupled, $d\varepsilon_{ij}^{\text{(c)}}$
                & 0.37$\times$10$^{-7}$
                & 0.37$\times$10$^{-7}$\\
    \midrule
    Total, $d\varepsilon_{ij}$
                & $-$20.00$\times$10$^{-7}$
                & 2.31$\times$10$^{-7}$\\
    \bottomrule
  \end{tabular}
\end{table}
The table reveals a slight difference between the reversible
and elastic increments and between the irreversible and
plastic increments, and this difference is the
coupled part in Eqs.~(\ref{eq:rcp}) and~(\ref{eq:parts}).
Reversible and recoverable (elastic)
strain increments are not equivalent,
as a small coupled part $d\varepsilon_{ij}^{\text{(c)}}$
accompanies loading.
Although the coupled axial
strain part $d\varepsilon_{11}^{\text{(c)}}$ may seem small,
the coupled part of volume change
$dv^{\text{(c)}}$ was larger when
compared with its elastic and plastic parts.
These results contrast with those of Wan and Pinheiro \cite{Wan:2014a},
who found no difference between plastic and irreversible increments of
strain
in their simulations of triaxial compression with sphere assemblies.
Our result is not unexpected, however,
since frictional contact dissipation,
an irreversible process that can contribute to the total
strain increment,
was found to occur during loading,
during unloading, and during strain-reversal,
indicating that the unloading increment is not entirely irreversible.
In Section~\ref{sec:multi}, we will see that
the small elastic-plastic coupling revealed in Table~\ref{table:epirc}
is for a probe direction that \emph{minimizes} such coupling,
and more significant coupling occurs in other
loading directions.
Also note that the reversible, coupled, and plastic
parts are not aligned,
but rather occur in different directions
with different relative rates of dilation,
a possibility suggested by Hueckel \cite{Hueckel:1976a}
and explored for other loading directions
(Section~\ref{sec:multi}).
\subsection{Stress probes with linear-frictional contacts}\label{sec:linfricprobe}
Similar results were also obtained in simulations using the simpler
linear-frictional contact model
(see Figs.~\ref{fig:cones}a and~\ref{fig:cones}b).
Table~\ref{table:LFtransitions} gives results
for the initial probe state
$\varepsilon_{11}=-$0.3\%,
which is during the strain-hardening phase of loading,
when the deviator stress had reached about 50\% of the peak strength
(Fig.~\ref{fig:StressStrain}).
The table
gives the fraction of elastic and slip
contacts during the tiny loading and retrogression probes
(again, a $d\varepsilon_{11}$ of $-2\times$10$^{-6}$).
\begin{table}
  \centering\small
  \caption{Results of loading and retrogression (strain-reversal)
           Type~1 probes, starting at strain
           $\varepsilon_{11}=-0.3\%$ with linear-frictional contacts.
           Transitions across the two contact conditions
           (elastic and slip) are shown between
           the loading (AB) and strain-reversal (AC) probes.
           \label{table:LFtransitions}}
  \begin{tabular}{ll|cc|c}
    \toprule
    & &\multicolumn{2}{c|}{Retrogression} & Loading\\
    \cline{3-4}
    & & \rule{0ex}{2ex}Elastic & Slip & totals\\
    \midrule
    \multicolumn{1}{l|}{Loading} & Elastic  & 0.878 & 0.000 & 0.878 \\
    \multicolumn{1}{l|}{\ }       & Slip    & 0.111 & 0.011 & 0.122 \\
    \midrule
    \multicolumn{2}{l}{Retrogression totals} & 0.989 & 0.011 & 1.000 \\
    \bottomrule
  \end{tabular}
\end{table}
Similar to Fig.~\ref{fig:Probe1}c,
the table presents these fractions of contacts in terms
of the transitions of elastic and slip modes
between loading and retrogression
(strain-reversal).
Only four transitions are applicable between the two modes,
and three of these occurred.
For example,
of the 11.1\% of contact that were sliding during loading,
about 10\% ($=1.1\%$$/$11.1\%) continued to slide during
the strain-reversal probe.
As with the more complex Cattaneo--Mindlin contacts,
some contacts continued to slip during a reversal of strain.
\par
Types~2 and~3 probes were also performed
with linear-frictional contacts.
The results of
loading and retrogression (Type~2) probes
are shown in Table~\ref{table:ElPlProbe2}.
\begin{table}
  \caption{Results of loading and reversible
           (Type~2) probes, starting at strain
           $\varepsilon_{11}=-0.3\%$ with
           linear-frictional contacts.
           Reversible strains were determined
           by artificially suppressing contact sliding.
           \label{table:ElPlProbe2}}
  \centering\small
  \begin{tabular}{lcrrrr}
    \toprule
        &  &
          \multicolumn{1}{c}{Reversible}
        & \multicolumn{1}{c}{Irreversible}
        & \multicolumn{1}{c}{Coupled}
        & \multicolumn{1}{c}{Total}\\
    Probes & Strain &
          \multicolumn{1}{c}{$d\varepsilon^{\text{(r)}}_{ij}$}
        & \multicolumn{1}{c}{$d\varepsilon^{\text{(i)}}_{ij}$}
        & \multicolumn{1}{c}{$d\varepsilon^{\text{(c)}}_{ij}$}
        & \multicolumn{1}{c}{$d\varepsilon_{ij}=d\varepsilon^{\text{(r)}}_{ij}+d\varepsilon^{\text{(i)}}_{ij}$}\\
    \midrule
    Types 2 \& 3 & $d\varepsilon_{11}$ &
             $-$4.21$\times$10$^{-7}$ & $-$15.79$\times$10$^{-7}$ &
             0.43$\times$10$^{-7}$
             & $-$20.00$\times$10$^{-7}$\\
            & $dv$ &
            1.56$\times$10$^{-7}$ & $-$0.71$\times$10$^{-7}$ &
            0.33$\times$10$^{-7}$
            & 0.85$\times$10$^{-7}$\\
    \bottomrule
  \end{tabular}
\end{table}
The results are
similar to those with Cattaneo--Mindlin contacts
(Section~\ref{sec:Type2CM}):
about 79\% of the loading increment $d\varepsilon_{11}$
is irreversible;
the reversible part of volume change was dilative;
and the irreversible part of volume change was contractive
(see Section~\ref{sec:progress} for further discussion).
With loading-unloading (Type~3) probes,
also included in Table~\ref{table:ElPlProbe2},
the directions of the elastic and plastic increments
were the same as those with Cattaneo--Mindlin contacts
(as in Table~\ref{table:epirc}), and
small coupled strain increments accompany
the closed path in stress-space.
%
\section{Multi-directional elastic-plastic coupling}\label{sec:multi}
The results discussed above are exclusively
for axisymmetric triaxial conditions,
in which the initial probe state was established with constant-$p$
triaxial compression ($\varepsilon_{11}<0$) and
in which the subsequent triaxial probes
were conducted with equal lateral strains,
$d\varepsilon_{22}=d\varepsilon_{33}$.
We now consider multi-directional (i.e., true-triaxial) probes that
began from initial probe states that were established,
as in the previous section,
with constant-$p$ triaxial compression.
One set of probes was conducted within the
deviatoric pi-plane of isochoric or isobaric
strain increments;
another set of probes was of generalized triaxial loading
within the Rendulic plane of volumetric and deviatoric strain increments.
With each series of probes (in the pi-plane or Rendulic plane),
we conducted over 50 probes in different directions,
and both
Types~2 and~3 probes were used to measure the elastic, plastic,
and coupled parts of the total strain increment for each probe direction.
\par
\begin{figure}
  \centering
  \includegraphics{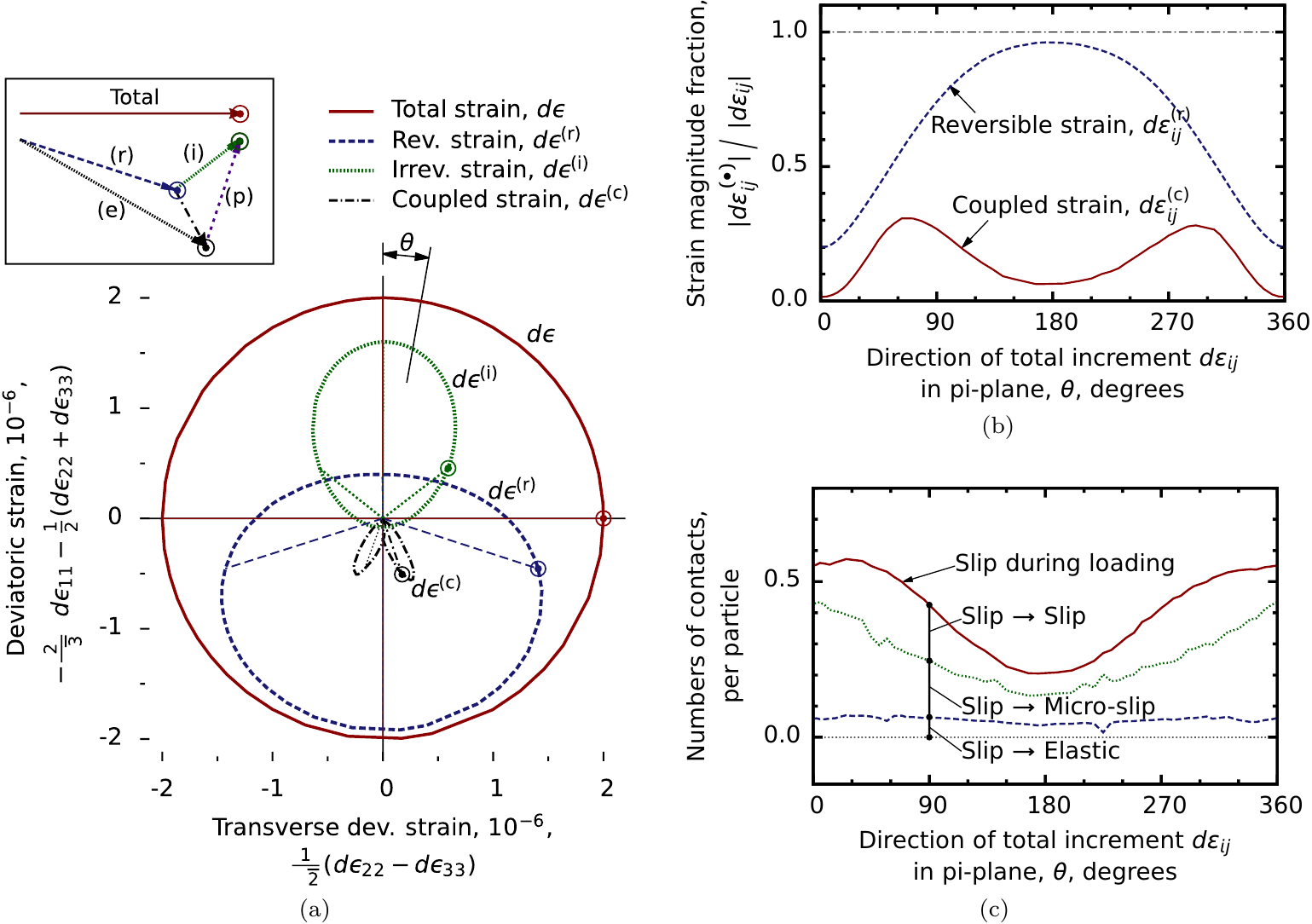}
  \caption{Isochoric probes of magnitude
           $|d\varepsilon_{ij}|=2\times 10^{-6}$
           at strain $\varepsilon_{11}=-0.3\%$ for the
           assembly with Cattaneo--Mindlin contacts:
           (a)~pi-plane of deviatoric and transverse-deviatoric
           strain increments;
           (b)~magnitudes of the reversible and coupled strain increments
           relative to the magnitude of the total strain increment;
           and (c)~transitions for the loading and unloading
           probes among the set of contacts that were slipping during
           loading.
           Dots ``$\odot$'' are for a total increment
           $d\varepsilon_{ij}$ of transverse deviatoric loading
           ($\theta=90^{\circ}$),
           and the inset at the top of Fig.~\ref{fig:piplane1}a
           shows the addition of the
           $d\varepsilon_{ij}^{\text{(r)}}$ and
           $d\varepsilon_{ij}^{\text{(i)}}$ vectors.
           \label{fig:piplane1}}
\end{figure}

Figure~\ref{fig:piplane1} shows
strain results within the pi-plane for isochoric (undrained) probes
that were initiated at strain $-\varepsilon_{11}=0.3\%$ for the
assembly with Cattaneo-Mindlin contacts.
These results are for the early stage of strain hardening at
which the material had reached 55\% of the peak strength.
Again,
this initial probe state was reached with constant-$p$ triaxial
compression (see Fig.~\ref{fig:StressStrain}).
Probes of \emph{equal strain-increment magnitude} were used
for each loading direction, with
$|d\boldsymbol{\varepsilon}_{ij}|=
 (d\varepsilon_{11}^{2}+d\varepsilon_{22}^{2}+d\varepsilon_{33}^{2})^{1/2}
 =2\!\times\!10^{-6}$.
The pi-plane strains in Fig.~\ref{fig:piplane1}a have been
scaled so that strain increments of equal magnitude
are equidistant from the origin.
The upward vertical direction, $\theta=0^{\circ}$, is for
continued isochoric triaxial compression
($d\varepsilon_{11}<0$,
$d\varepsilon_{22}=d\varepsilon_{33}=-\tfrac{1}{2}d\varepsilon_{11}$);
and
the downward direction, $\theta=180^{\circ}$, is for triaxial extension
($d\varepsilon_{11}>0$,
$d\varepsilon_{22}=d\varepsilon_{33}=-\tfrac{1}{2}\varepsilon_{11}$).
The horizontal directions, $\theta=90^{\circ}$ and
$270^{\circ}$, correspond to transverse deviatoric loading
($d\varepsilon_{11}=0$, $d\varepsilon_{22}=-d\varepsilon_{33}$).
\par
Each closed line in Fig.~\ref{fig:piplane1}
is a locus of over 50 strain increments~---
lines of total
$d\varepsilon_{ij}$, reversible $d\varepsilon^{\text{(r)}}_{ij}$,
irreversible $d\varepsilon^{\text{(i)}}_{ij}$, and
coupled $d\varepsilon^{\text{(c)}}_{ij}$~---
for different isochoric combinations (directions)
of the strain increments
$d\varepsilon_{11}$, $d\varepsilon_{22}$, and $d\varepsilon_{33}$.
During the stage of early strain hardening
that is represented in the figure, the reversible and irreversible
parts are of similar magnitude.
As an example, the ``$\odot$'' symbols
in Fig.~\ref{fig:piplane1}a show the incremental
parts of a total strain increment of transverse deviatoric
loading
(i.e., for a total increment $d\varepsilon_{ij}$ have
$\theta=90^{\circ}$,
which is an increment that
is orthogonal to both the volumetric and deviatoric directions).
For this particular loading direction,
the inset in the top left corner
of Fig.~\ref{fig:piplane1}a shows how the various parts combine to form
the total increment $d\varepsilon_{ij}$ (the small vector
in the middle of the inset is the coupled part
$d\varepsilon_{ij}^{\text{(c)}}$).
The inset shows the distinctly different directions of the
reversible, irreversible, elastic, plastic, and coupled
parts.
In the main plot of Fig.~\ref{fig:piplane1}a,
the locus of coupled strain increments appears as two lobes below
the horizontal axis, corresponding to increments roughly in
the direction of triaxial extension.
Not only is coupling present, but
the magnitude $|d\varepsilon_{ij}^{\text{(c)}}|$ is,
for some loading directions,
significant when compared with the reversible and irreversible
parts.
For certain loading directions, the coupled magnitude
is about 30\%
of the total increment $|d\varepsilon_{ij}|$.
We also note that in no deviatoric loading direction is the irreversible
increment (or, for that matter, the plastic increment) zero,
as the $d\varepsilon_{ij}^{\text{(i)}}$ locus fully envelops
the origin.
\par
The significance of elastic-plastic coupling during deviatoric
(pi-plane) loading is also shown in Fig.~\ref{fig:piplane1}b, which gives
the fractions
$|d\boldsymbol{\varepsilon}^{\text{(c)}}_{ij}|/|d\boldsymbol{\varepsilon}_{ij}|$
and
$|d\boldsymbol{\varepsilon}^{\text{(r)}}_{ij}|/|d\boldsymbol{\varepsilon}_{ij}|$
of the magnitudes of
coupled and reversible parts relative to the total
strain increment.
The results for different directions $\theta$
of the total increment $d\varepsilon_{ij}$
demonstrate that elastic strains are not entirely
reversible,
particularly for probe directions that are oblique to the
initial loading direction ($\theta=0^{\circ}$),
as these oblique directions produce significant coupled increments.
To place the results of Section~\ref{sec:probes} in the context
of those in Fig.~\ref{fig:piplane1},
the probes of the previous section were strictly
of triaxial compression
(i.e., in the upward direction, $\theta=0^{\circ}$,
of Fig.~\ref{fig:piplane1}a).
Of the $360^{\circ}$ of loading directions, the coupled increments
are \emph{smallest} for probes that continue in this original
(upward)
direction of loading.
\par
The micro-scale origin of elastic-plastic coupling is
the subject of
Fig.~\ref{fig:piplane1}c,
which quantifies the transitions of the contacts' conditions
for the loading and unloading phases of a closed
Type~2 path in stress-space.
In this figure, we consider only the subset of contacts
that were fully sliding during the loading phase of
isochoric probes in multiple directions
(i.e., for $d\varepsilon_{ij}$
angles $\theta$ of 0$^{\circ}$ to 360$^{\circ}$).
The top line give the numbers of sliding contacts during
loading (i.e., during the segment AB in Fig.~\ref{fig:probes}),
normalized by dividing by the number of particles.
This subset of contacts can transition to three different
conditions during the unloading phase
(when $d\sigma_{ij}\rightarrow -d\sigma_{ij}$, as in
segment BA$^\prime$ in Fig.~\ref{fig:probes}):
to an elastic condition, to micro-slip, or to continued
sliding (recall the ``slip'' row in Fig.~\ref{fig:Probe1}c).
Some fraction of contacts slide during loading,
regardless of the loading direction $d\varepsilon_{ij}$.
Moreover, some of the contacts that were sliding during loading
continue to slide during unloading.
For example, when the loading increment was in
the direction triaxial compression
($\theta=0^{\circ}$), 0.56 contacts per particle were slipping.
During the subsequent unloading increment, $0.56-0.42=0.14$
continued to slip;
$0.42-0.06=0.36$ transitioned to micro-slip;
and 0.06 contacts per particle had transitioned to an elastic condition.
\par
\begin{figure}
  \centering
  \includegraphics{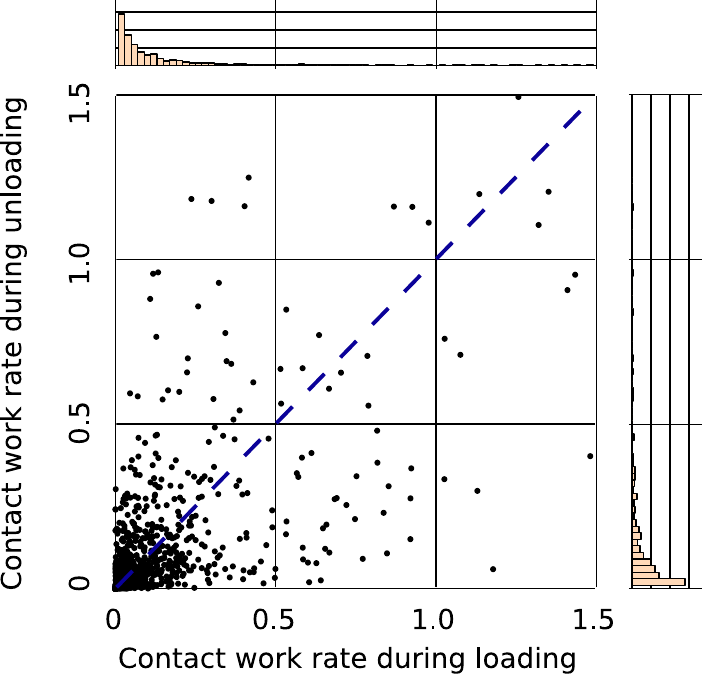}
  \caption{Comparison of sliding rates (work rates)
           among contacts
           during the loading and unloading phases of a single
           isochoric probe with $\theta=90^{\circ}$.
           The probe was performed at strain
           $\varepsilon=-0.3\%$ with Cattaneo--Mindlin contacts.
           \label{fig:workrates}}
\end{figure}

In Fig.~\ref{fig:workrates}, we show the vigor of contact sliding
for a single isochoric loading--unloading probe: a probe in the
direction $\theta=90^{\circ}$ of $d\varepsilon_{ij}$, which
corresponds to deviatoric loading that is transverse to the
original triaxial compression
(refer to the ``$\odot$'' points in Fig.~\ref{fig:piplane1}a).
The figure represents
those contacts that were sliding during both loading
and unloading phases of the probe
(refer to the ``slip$\rightarrow$slip'' line
in Fig.~\ref{fig:piplane1}c),
and each dot represents a single
contact among the 1980 contacts that were sliding
during both loading and unloading
(about 4530 contacts were sliding during loading,
but sliding had ceased at all but the 1980 contacts during
the unloading phase).
The vigor of sliding is given as the rate of frictional
dissipation at each contact, with a contact's work rate
normalized by dividing by $p$,
by the cubed mean particle size, and by the bulk strain rate.
Although fewer contacts are sliding during unloading,
sliding remains vigorous during unloading,
and among many contacts, frictional dissipation was
\emph{greater} during unloading than during loading.
These micro-scale movements are consistent with
the results in Fig.~\ref{fig:Rose}, in which the directions
of contact movement during loading and retrogression
(strain reversal) are only weakly counter-correlated,
and many contacts continue to move and slip in the same direction
during both loading and its reversal.
\par
\begin{figure}
  \centering
  \includegraphics{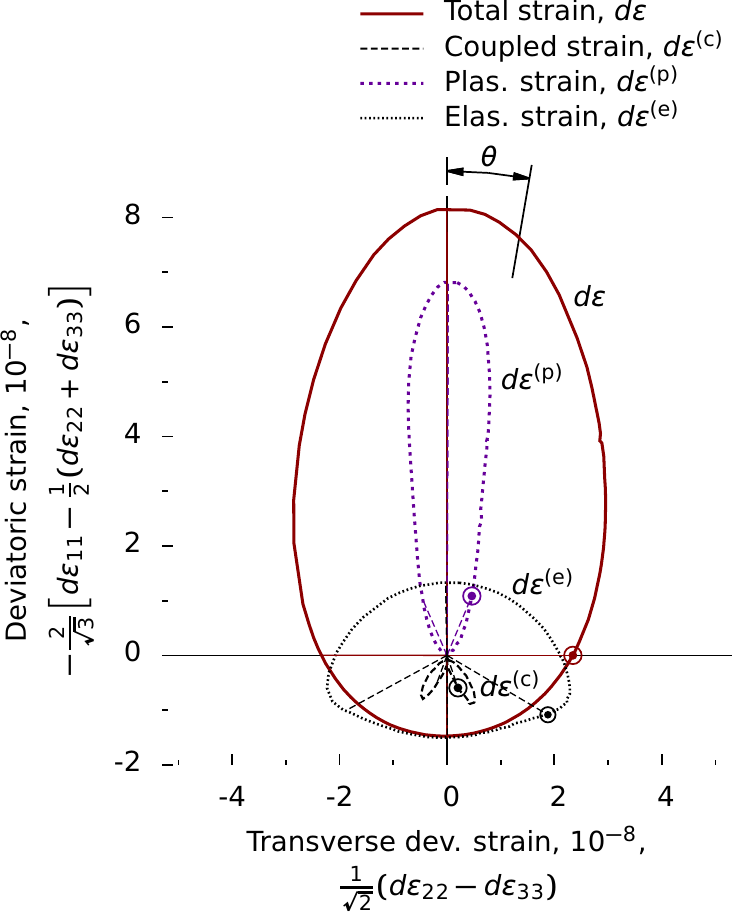}
  \caption{Isobaric probes of magnitude $|d\sigma_{ij}|=1$~Pa:
           total, elastic, plastic, and coupled parts of
           strain increments
           for an assembly with Cattaneo--Mindlin contacts
           at strain $\varepsilon=-0.3\%$.
           Dots ``$\odot$'' are for a total increment
           $d\varepsilon_{ij}$ of transverse deviatoric loading
           ($\theta=90^{\circ}$),
           and the inset at the top of Fig.~\ref{fig:piplane1}a
           shows the addition of the
           $d\varepsilon_{ij}^{\text{(e)}}$ and
           $d\varepsilon_{ij}^{\text{(p)}}$ vectors.
           \label{fig:piplane2}}
\end{figure}

Figure~\ref{fig:piplane2}
shows the results of deviatoric, pi-plane
isobaric (constant-$p$)
probes of \emph{equal stress magnitude},
$|d\sigma_{ij}|=1$~Pa for the assembly with
Cattaneo--Mindlin contacts.
Each probe had a strain magnitude of $2\times 10^{-6}$,
and after noting the resulting stress increment,
the strains were scaled to a stress magnitude of 1~Pa.
The loci of total, elastic, plastic, and coupled strain increments
are displayed as ``response envelopes,''
in the manner of Gudehus \cite{Gudehus:1979a}.
The effect of elastic-plastic coupling is clear:
contrary to conventional models,
the elastic strain increment is not of a linear form
$C_{ijkl}\,d\sigma_{kl}$
(note the \emph{sum} of the first two terms on the
right of Eq.~\ref{eq:depsep}),
as the
envelope of elastic increments
$d\varepsilon_{ij}^{\text{(e)}}$
is an odd shape and is
certainly not an ellipse that would fit
the linear form.
This result is cautionary: the recoverable (elastic) strain increment
is not a linear function of the stress increment, as in
a generalized Hooke's law.
One must recall that
the elastic moduli are altered by irreversible processes
(second term on the right of Eq.~\ref{eq:depsep}),
as particle rearrangements and frictional contact
dissipation in the form
of slip and micro-slip accompany plastic deformation.
On the other hand,
in a separate analysis of the \emph{reversible} strains,
we found that the
reversible strain increments, indeed, conform closely to
a linear form $C_{ijkl}\,d\sigma_{kl}$.
(see \cite{KuhnDaouadji:probes}).
\par
Note also that the plastic envelope is not a straight line emanating from
the origin in a single flow direction.
That is, plastic strains are not
of the form $g_{ij}f_{kl}\,d\sigma_{kl}$ for some yield
direction $f_{kl}$ and flow direction $g_{ij}$.
This aspect of the incremental behavior was noted
in the simulations of Kishino \cite{Kishino:2003a} and
Wan and Pinheiro \cite{Wan:2014a}, is quantified by
the authors in \cite{KuhnDaouadji:probes},
and is addressed by
certain constitutive models with incremental
non-linearity \cite{Darve:1974a,Darve:1982b,Dafalias:2016a}.
Finally, we note that the locus of plastic increments
$d\varepsilon_{ij}^{\text{(p)}}$ does not pass through the origin,
as the locus of plastic increments $d\varepsilon_{ij}^{\text{(p)}}$
envelops the origin,
and some plastic deformation occurs with all loading directions
(as with vanishing elastic range plasticity models \cite{Dafalias:1975c}).
%
\par
\begin{figure}
  \centering
  \includegraphics{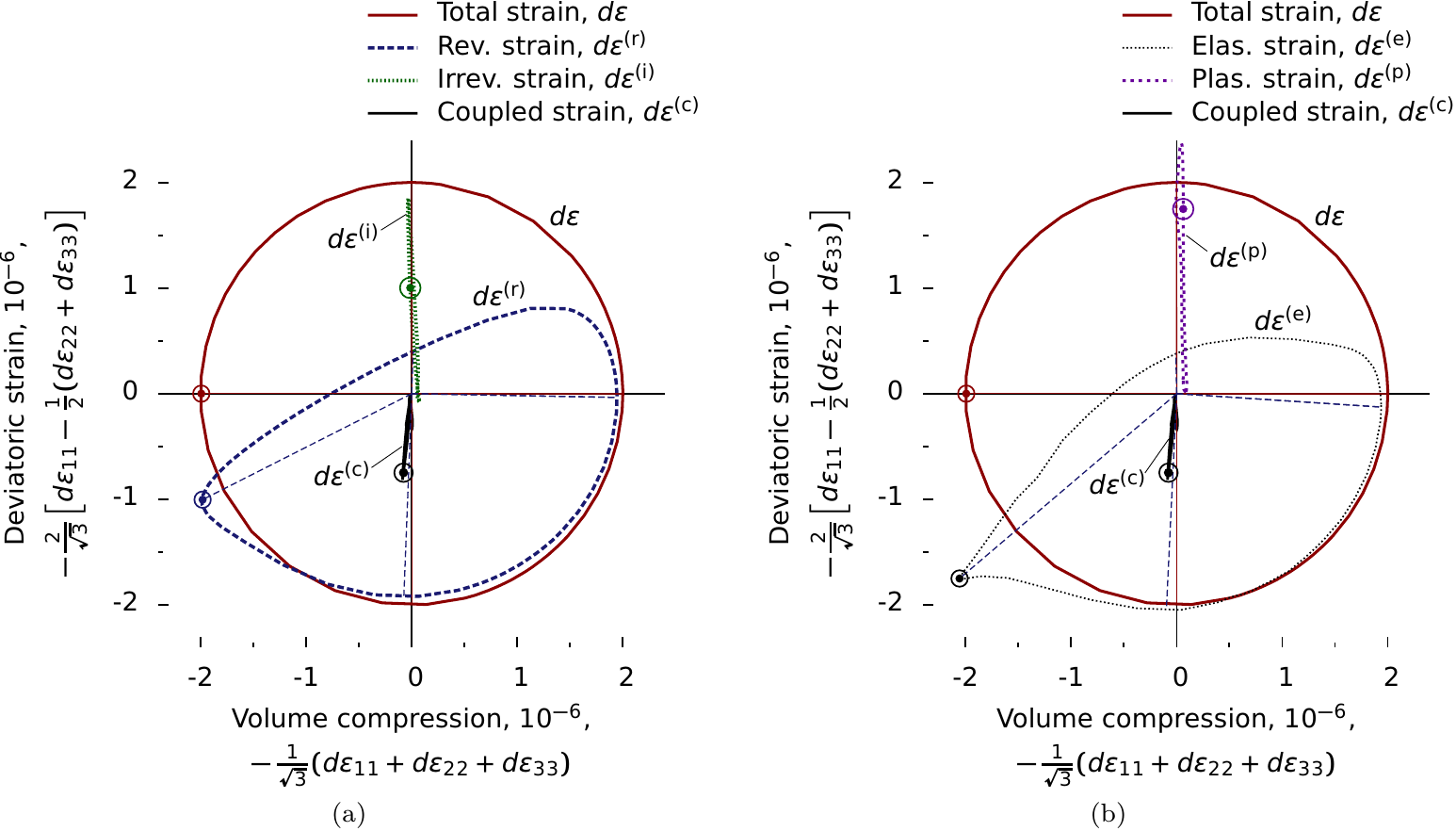}
  \caption{Probes within the triaxial Rendulic plane
           for the assembly with Cattaneo--Mindlin contacts
           at strain $\varepsilon=-0.3\%$:
           (a)~reversible and irreversible parts of the
           total strain increment; and
           (b)~elastic and plastic parts of the total strain increment.
           Dots ``$\odot$'' are for a total strain increment
           of isotropic, volumetric expansion.
           \label{fig:Rendulic}}
\end{figure}

A separate series of over fifty probes was conducted with different
combinations of deviatoric and volumetric strain increments under
axisymmetric triaxial conditions.
The results in Fig.~\ref{fig:Rendulic} are presented in the
Rendulic plane:
the horizontal axis is the volumetric direction
(with compression to the right), and
the vertical axis is deviatoric strain,
as with the vertical axes in Figs.~\ref{fig:piplane1}a
and~\ref{fig:piplane2} (upward represents continued
axial compression).
As with the earlier figures, the probes
in Fig.~\ref{fig:Rendulic} are of magnitude
$|d\varepsilon_{ij}|=2\times 10^{-6}$ and
were begun during strain hardening, when
the stress had reached 55\% of peak strength
($\varepsilon_{11}=-0.3\%$, Fig.~\ref{fig:StressStrain}).
As an example, the dots ``$\odot$'' show the various incremental
parts for a total isotropic increment of volume expansion.
The coupled increments are seen to occur primarily in the
negative deviatoric direction (i.e. triaxial extension)
and can be fairly
large when compared with the total strain increment.
The locus of irreversible parts is a nearly vertical line,
indicating the flow is almost exclusively deviatoric
(the initial probe strain of 0.3\% is near the
transition between contractive and dilatant behaviors).
A comparison of the reversible and elastic increments
in Figs.~\ref{fig:Rendulic}a and~\ref{fig:Rendulic}b reveals
a sharp, non-smooth shape of the elastic response,
a consequence of adding the spiked locus of coupled
increments to the smooth locus of reversible increments.
As was noted with behavior in the pi-plane,
the coupled strain part is smallest
when the total strain increment is
in the direction of continued triaxial compression,
but for other directions, the coupled part can
be as large as 30\% of the total increment.
The coupled part is largest during volumetric
expansion and is primarily deviatoric.
\par
The simulation results in this section are for the assembly
with Cattaneo--Mindlin contacts at the initial probes
state of strain $\varepsilon_{11}=-0.3\%$.
We conducted similar simulations with the assembly
having linear--frictional contacts,
and the results were qualitatively similar in every respect.
Results at other strains are reviewed in the next section.
\section{Evolution of reversible and irreversible increments}%
\begin{figure}
  \centering
  \includegraphics{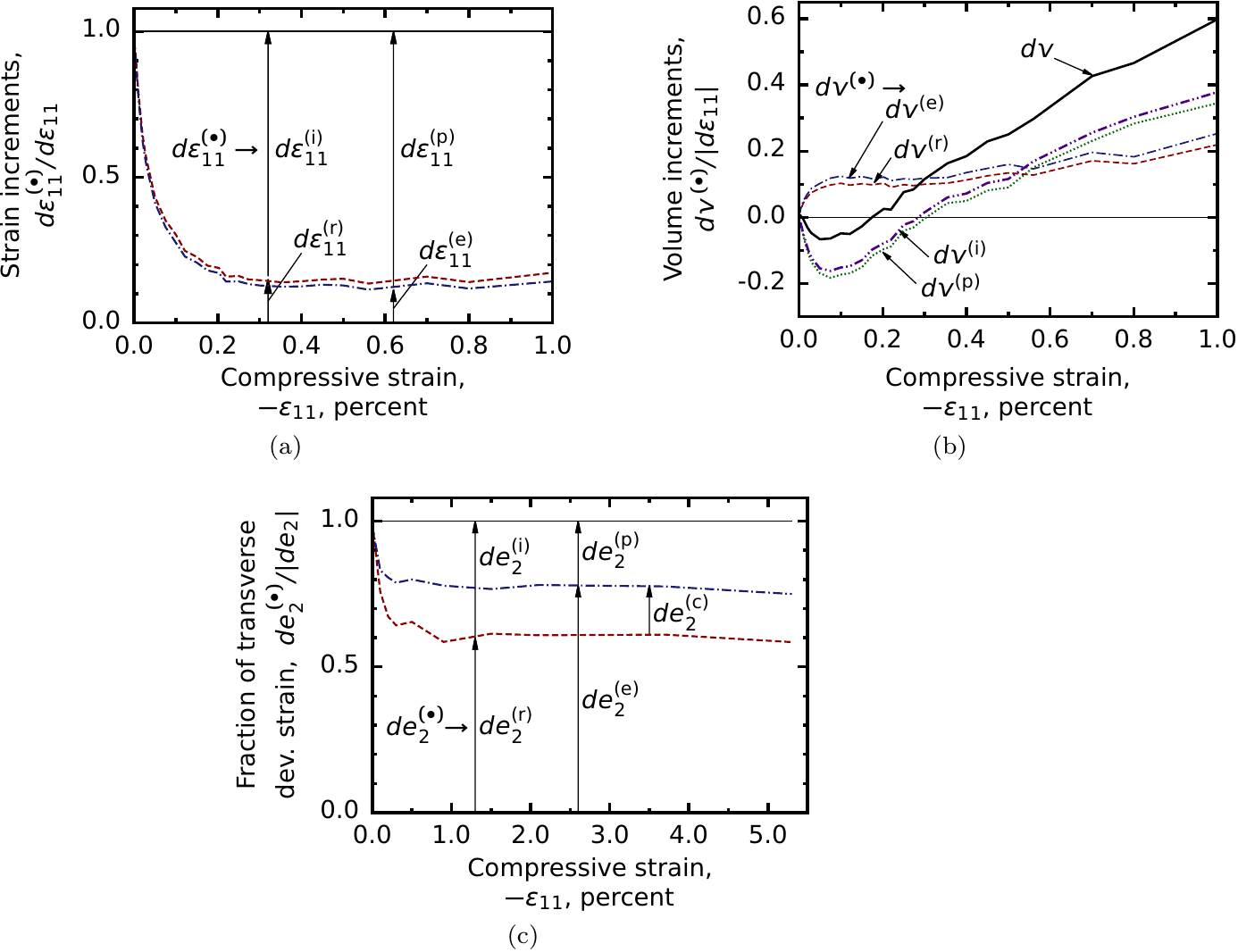}
  \caption{Reversible--irreversible and elastic--plastic
           partitions of
           strains during constant-$p$ triaxial compression
           of the assembly with Cattaneo--Mindlin contacts:
           (a)~axial increments $d\varepsilon_{11}$ for
           probes in direction $\theta=0^{\circ}$
           (see Fig.~\ref{fig:piplane1}a);
           (b)~volumetric increments $dv$ for strains
           in direction $\theta=0^{\circ}$; and
           (c)~transverse-deviatoric strain in
           direction $\theta=90^{\circ}$.
           Strain increments are normalized by dividing by
           the increments $d\varepsilon_{11}$
           and $de_{2}$.
           The superscript $(\odot)$ in the
           axes labels can mean (e), (p), (i), (r),
           or (c).
           \label{fig:progress}}
\end{figure}

\label{sec:progress}
We now consider the evolution of the reversible, irreversible,
and coupled parts of a strain increment, from the start
of triaxial loading until a compressive strain
of 1\% and larger.
We conducted Types~2 and~3
stress probes, similar to those described
above, at several stages of the
constant-$p$ triaxial compression
(as in Fig.~\ref{fig:StressStrain}, with axial
strain $\varepsilon_{11}$ at constant mean stress $p$).
Figures~\ref{fig:progress}a and~\ref{fig:progress}b
show the reversible--irreversible and elastic--plastic
partitions of the axial and volumetric strain
increments for the assembly
with Cattaneo--Mindlin contacts,
in which probes of the
total strain $d\varepsilon_{ij}$
were conducted in the direction
of continued axisymmetric
constant-$p$ triaxial compression
(direction $\theta=0^{\circ}$ in Fig.~\ref{fig:piplane1}a).
Figure~\ref{fig:progress}c shows these partitions
for probes conducted in the transverse-deviatoric direction
(direction $\theta=90^{\circ}$ in Fig.~\ref{fig:piplane1}a).
Here, $de_2$ represents the transverse deviatoric strain
increment $(d\varepsilon_{22}-d\varepsilon_{33})/\sqrt{2}$,
which is measured along the horizontal axis in
Figs.~\ref{fig:piplane1} and~\ref{fig:piplane2}.
The various incremental parts are normalized by dividing by the
full increments, either $d\varepsilon_{11}$
or $de_{2}$.
In Figs.~\ref{fig:progress}a and \ref{fig:progress}c,
the strain partitions are shown in an additive manner;
whereas, for the volume increments
of Fig.~\ref{fig:progress}b, the four parts
(elastic, plastic, reversible, and irreversible) are shown
separately.
\par
In regard to deviatoric probes in direction
$\theta=0^{\circ}$,
the behavior was almost entirely reversible (and elastic)
at the start of loading,
with irreversible increments
that were nearly zero.
Irreversible (and plastic) strain increments increased with increasing
axial deformation and dominated the loading behavior at
cumulative strains greater than 0.1\%.
Volume change was contractive at the start of loading and transitioned
to dilative behavior at strain 0.2\% (Fig.~\ref{fig:progress}b,
note that the transition for plastic strains occurred at
a strain of about 0.3\%).
Throughout the constant-$p$ loading (i.e. axial loading while
maintaining constant mean stress),
the reversible (and elastic) deformation was dilative with
a rate $dv^{\text{(r)}}/|d\varepsilon_{11}|$ between 0 and 0.23.
The particle arrangement was initially isotropic, and at zero strain,
the elastic response exhibited no volume change during the constant-$p$
loading.
Upon further loading, however,
the fabric became increasingly anisotropic,
as the material became elastically
stiffer in the axial $d\varepsilon_{11}$ direction
(elastic hardening), while becoming softer in the transverse
directions (elastic softening).
At a strain of 0.5\%, the reversible Young's modulus
in the axial direction was more than twice the transverse modulus,
giving rise to the reversible dilation, for constant-$p$ loading, seen in
Fig.~\ref{fig:progress}b
(see \cite{KuhnDaouadji:probes}
for further analysis of the reversible moduli).
The irreversible (and plastic) volume rate transitioned
from contractive, at the start of loading,
to dilative, at strains greater than 0.2\%
and was responsible for the initial contractive
behavior.
Throughout loading, a
coupled ``c'' part was present, although small,
for the probes in direction $\theta=0^{\circ}$
of continued triaxial loading,
and the coupled part
appears as the thin difference between the elastic
(``e'') and recoverable (``r'') lines.
\par
Although it is small for probes of continued triaxial
compression,
Fig.~\ref{fig:progress}c shows that the
coupled part was significant for transverse-deviatoric
probes in direction $\theta=90^{\circ}$
(see Figs.~\ref{fig:piplane1} and~\ref{fig:piplane2}
and the ``$\odot$'' symbols that correspond to the
$\theta=90^{\circ}$ direction).
The coupled part of the strain
increment, $de_{2}^{\text{(c)}}$, is significant
for this probe direction and accounts for 15--20\%
of the total transverse strain for axial strains
greater than 0.2\%.
\section{Conclusions}
Bulk deformation of a granular material
produces contact frictional dissipation in the form of
slip and micro-slip at the inter-particle contacts.
The aggregate
effect of this contact-scale dissipation is stress-work expended
in plastic deformation and in irreversible alterations
of the particle arrangements and contact conditions.
When a strain increment is reversed,
$d\boldsymbol{\varepsilon}\rightarrow -d\boldsymbol{\varepsilon}$,
or when a stress increment is reversed,
$d\boldsymbol{\sigma}\rightarrow -d\boldsymbol{\sigma}$,
the contact movements are not necessarily reversed, and
some contacts that were undergoing slip or micro-slip during a
forward increment continue
to slip during the reversed retrogression or unloading increment.
That is,
frictional contact dissipation and irreversible deformation
occur during both forward and reversed increments,
and if a strictly elastic region of loading exists at all, it is smaller
than the strain increments of
$2\times 10^{-6}$ that were used in our simulations.
In the context of thermomechanics,
granular materials are coupled materials in the sense that
increments of elastic strain are altered by changes in the elastic
moduli that occur concurrently with the loading increment
and are associated with irreversible micro-scale
changes within the material.
The difference between the conventional elastic (recoverable)
strain increment and the reversible part
of the increment is a coupling part
$d\boldsymbol{\varepsilon}^{\text{(c)}}$.
This coupled part is small during conventional triaxial
loading conditions and might escape notice in a laboratory setting
or even in DEM simulations, but the
coupled increments are quite large for other loading directions
and produce a significant difference between the elastic--plastic
and reversible--irreversible partitions
of a total strain increment.
\par
Our measurements of the
reversible, irreversible, and coupled parts
of strain increments
were obtained through rather extraordinary intervention:
by suppressing the slip and micro-slip that would otherwise
occur at the contacts, as in \cite{Calvetti:2003a}.
This sort of intrusion or internal constraint
is not available with laboratory testing,
which relies on loading--unloading cycles to determine the elastic and
plastic parts of strain.
Even though our simulation methods
can not be replicated in physical tests,
the implications of our simulation are certainly cautionary:
elastic strain increments 
are accompanied by micro-scale irreversible processes
that alter the elastic moduli during plastic deformation,
and the application
of elastic strains in constitutive formulations should be done
with this understanding.
For example, the shape of the response envelopes of
elastic increments in
Fig.~\ref{fig:piplane2}
indicates the difficulty of applying a linear Hooke stiffness
to the
\emph{elastic} strain increments,
although in a related work we show that a linear stiffness relation
precisely fits the \emph{reversible} increments
\cite{KuhnDaouadji:probes}.
%
%
\par
All of the forgoing conclusions apply to simulations with
either of two contact models: with
the simple linear-frictional model and with the
more complex Hertz-type Cattaneo--Mindlin model.
Although we have described behavior that is complex
and inconsistent with simple constitutive models,
the simulations do show that the bulk large-strain behavior
of granular materials during monotonic loading is somewhat insensitive
to the details of the contact model, provided that the simulations
include a common friction coefficient.
%
%
\par
Our study was limited to behavior that followed
an initial monotonic
drained triaxial compression, and
this work could certainly be extended to more general loading paths:
other proportional loading paths, paths involving principal stress
rotations, and cyclic paths.
The results suggest, however, that granular behavior is even more complex
than previously thought and that granular materials should be
treated as coupled and thoroughly inelastic.
%
\pagebreak
\appendix
\section{Incremental behavior of contact models}\label{app:contacts}
\begin{figure}
  \centering
  \includegraphics{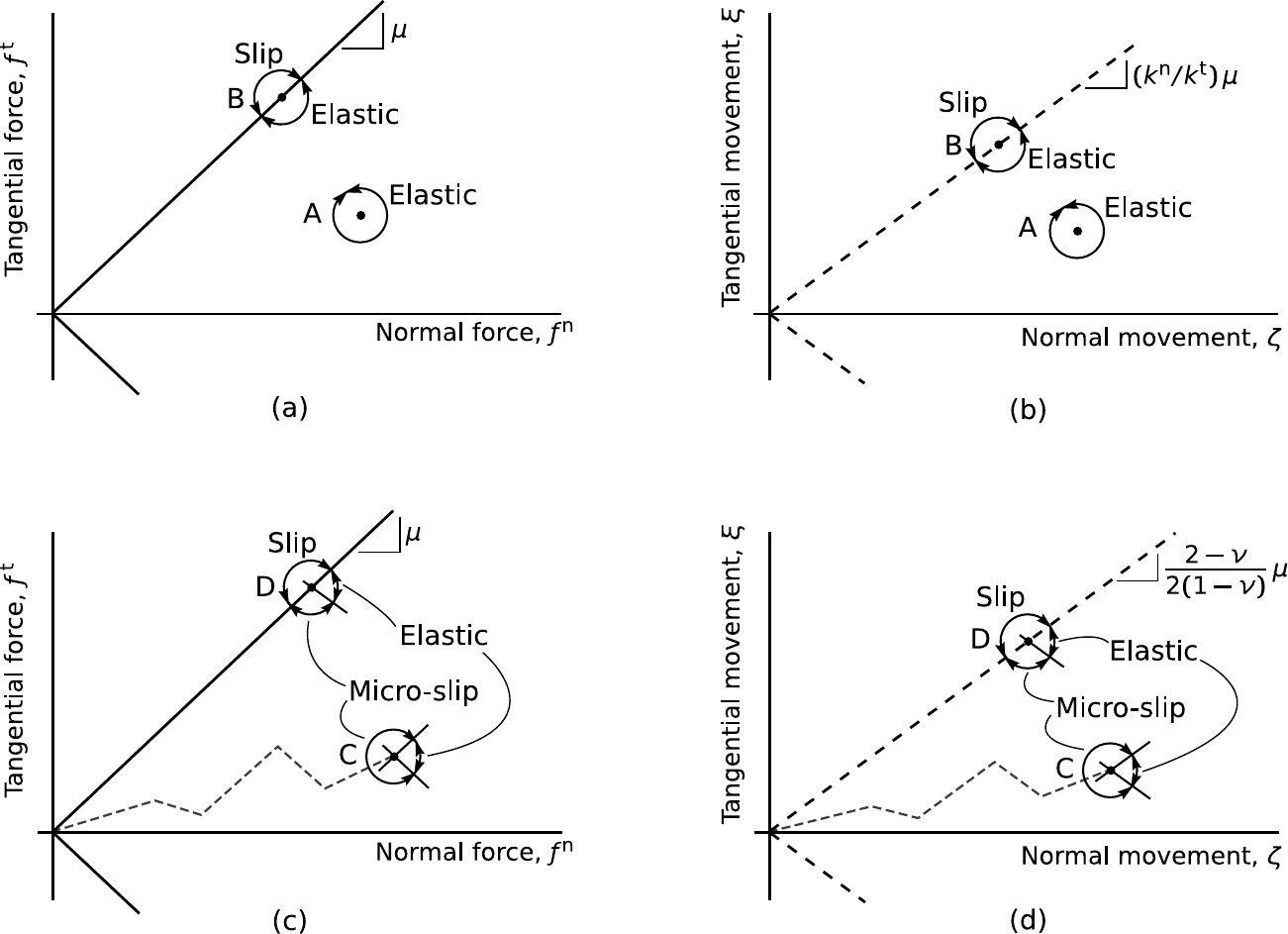} 
  \caption{Contact yield surfaces in force-space
           and movement-space for two contact models:
           (a)~force-space for
           linear-frictional contacts;
           (b)~displacement-space for linear-frictional contacts;
           (c)~force-space for Cattaneo--Mindlin contacts; and
           (d)~displacement-space
           for Cattaneo--Mindlin contacts.
           \label{fig:cones}}
\end{figure}

Figure~\ref{fig:cones} depicts the two contact models~--- linear-frictional
and Cattaneo--Mindlin~--- that were used in the simulations.
Figures~\ref{fig:cones}a and~\ref{fig:cones}b
are of the simpler linear-frictional model,
showing
the two-dimensional spaces of forces
and movements,
with $f^{\text{n}}$ and $\zeta$ representing the normal
components of compressive contact force and displacement,
and $f^{\text{t}}$ and $\xi$ representing the tangential
displacement and force.
(In a three-dimensional setting, $\mathbf{f}^{\text{t}}$ and
$\boldsymbol{\xi}$
are vectors that lie in the contact's tangent plane.)
The behavior of the linear-frictional model is entirely
reversible for points within the contact yield cone having its
apex at
the $f^{\text{n}}$--$\mathbf{f}^{\text{t}}$ origin
(e.g., the behavior is reversible at points ``A'' in
Figs.~\ref{fig:cones}a and~\ref{fig:cones}b).
When the frictional limit is reached (points ``B'') and
$|\mathbf{f}^{\text{t}}|=\mu f^{\text{n}}$,
slip can commence and
the zone of incremental
reversibility is a
semi-space inside of the yield cone.
In the force-space of Fig.~\ref{fig:cones}a,
the reversible and irreversible (``slip'')
conditions are represented by small arced arrows
(actually, filled cones in 3-space) that show the
corresponding ranges of the force increments.
In the strain-space of Fig.~\ref{fig:cones}b,
reversible and irreversible (slip) increments
are shown in a similar manner, but the orientation of the yield
boundary (cone) has slope $\mu k^{\text{n}}/k^{\text{t}}$, where
$k^{\text{n}}$ and $k^{\text{t}}$ are
the linear contact stiffnesses in the
normal and tangential directions.
The increment of
frictional energy dissipation during a small
tangential movement $\Delta\boldsymbol{\xi}$
is computed as the incremental
difference of the work done by the tangential force,
$\mathbf{f}^{\text{t}}\cdot\Delta\boldsymbol{\xi}$,
and the change of elastic energy of the contact,
$\Delta[(\mathbf{f}^{\,\text{t}})^{2}/2k^{\text{n}}]$.
\par
The simple linear-frictional
contact yields a small-strain bulk shear modulus
that is independent of the
mean stress~--- a condition that
poorly represents the behavior of most geomaterials.
For this reason, a generalized Cattaneo--Mindlin contact
is considered a better representation of the particle
interactions \cite{Kuhn:2014c}.
\par
The incremental force-displacement relation
of the Cattaneo--Mindlin contact
is 
quite complex and depends on the
history of both normal and tangential movements.
Mindlin and Deresiewicz \cite{Mindlin:1953a}
derived the contact stiffnesses for
eleven such histories of contact movement, and
J\"{a}ger \cite{Jager:2005a} developed a general framework for the
Cattaneo--Mindlin contact that encompasses arbitrary sequences
of contact movement.
The ``J\"{a}ger contact'' is
incorporated in the authors'
OVAL code (see \cite{Kuhn:2011a,Kuhn:2014c}).
Because contact behavior is history-dependent,
a contact's history is recorded as an
``equivalent load history'', represented by
the dashed lines in Figs.~\ref{fig:cones}c
and Figs.~\ref{fig:cones}d: a piece-wise linear
sequence of displacement pairs $\zeta^{*}$
and $\boldsymbol{\xi}^{*}$ (or, in force-space,
pairs $\mathbf{f}^{*\,\text{t}}$ and $f^{*\,\text{n}}$).
The full list of these pairs comprise the internal
condition (the ``internal variables'') of
this single contact.
For forces that lie \emph{inside} the cone of full frictional slip
(points ``C'', Fig.~\ref{fig:cones}),
the incremental reversible
domain is quite restricted.
Reversible contact displacement is limited
to the following:
(1) increments that lie
inside the cone of slope $\mu\kappa$ in displacement-space
($\mu$ in force-space)
and (2) paths of perfect
reversal that ``back track'' along the equivalent load history
(factor $\kappa=(2-\nu)/2(1-\nu)$, $\nu$
being the Poisson ratio $\nu$ of the
grains \cite{Jager:2005a,Kuhn:2011a}).
Dissipative
annular micro-slip occurs for all other
increments.
When the frictional limit is reached
(i.e., $|\mathbf{f}^{\text{t}}|=\mu f^{\text{n}}$,
as at points ``D''),
the incremental contact behavior can be one of three types:
reversible, micro-slip (reversible-irreversible),
or slip (irreversible),
depending on the loading direction.
Compared with the linear-frictional model of
Figs.~\ref{fig:cones}a and~\ref{fig:cones}b,
a Hertz-like Cattaneo--Mindlin
contact is more complex and has
smaller reversible zones (cones), 
and this
contact behavior is expected
to produce a smaller, more restricted reversible region for the
bulk material behavior than that of the simpler linear-frictional
model.
\par
The computation of frictional contact
dissipation is also straightforward
with a Cattaneo--Mindlin contact,
provided that the contact is in full frictional slip
(points ``D'' in Fig.~{\ref{fig:cones}}).
Computing
dissipation is more difficult in the case of
micro-slip for Cattaneo--Mindlin contacts
that have not yet reached the frictional limit
(points ``C'' in Fig.~{\ref{fig:cones}}). 
For each time increment $\Delta t$
in our DEM simulations with Cattaneo--Mindlin
contacts,
we computed the increment of reversible tangential displacement
$\Delta\boldsymbol{\xi}^{\text{(r)}}$,
subtracted this reversible increment from
the full increment $\Delta\boldsymbol{\xi}$ to
find the irreversible tangential increment,
\begin{equation}\label{eq:CMir}
  \Delta\boldsymbol{\xi}^{\text{(i)}} =
  \Delta\boldsymbol{\xi} -
  \Delta\boldsymbol{\xi}^{\text{(r)}}
\end{equation}
and then computed the inner produce
of the irreversible increment
and the tangential force to find the increment of
dissipation:
\begin{equation}\label{eq:ftDissipation}
  \text{Contact dissipation} =
  \mathbf{f}^{\text{t}} \cdot
  \Delta\boldsymbol{\xi}^{\text{(i)}}
\end{equation}
To find the incremental reversible tangential movement
$\Delta\boldsymbol{\xi}^{\text{(r)}}$ in
Eq.~(\ref{eq:CMir}),
which represents the response to the
increment of tangential force
$\Delta\mathbf{f}^{t}$
\emph{in the absence of micro-slip},
we used Walton's solution for the tangential
contact response of rough (zero-slip) spheres
\cite{Walton:1978a}:
\begin{equation}
  \Delta\boldsymbol{\xi}^{\text{(r)}}
  =
  \frac{2-\nu}{8G}\frac{1}{a}
  \Delta\mathbf{f}^{\text{t}}
\end{equation}
where $G$ is the particle's shear modulus,
$\nu$ is its Poisson ratio, and $a$ is the current
radius of the contact zone.
For sphere--sphere contact, $a$ is the geometric
mean of the contact indentation and sphere radius
($a=\sqrt{\zeta R}$); for non-spherical asperities
of the power-law type ($z=A_{\alpha}r^{\alpha}$),
J\"{a}ger \cite{Jager:1999a} derived the radius as
\begin{equation}
  a = \left[
      \frac{\zeta}{\alpha\sqrt{\pi}}
      \frac{\Gamma(\frac{\alpha+1}{2})}
           {\Gamma(\frac{\alpha+2}{2})}
      \right]^{1/\alpha}
\end{equation}
where $\Gamma(\cdot )$ is the gamma function.
\par
In this manner, we tracked the frictional dissipation of each
Cattaneo--Mindlin contact so that
the total contact frictional
dissipation and total contact elastic energy of the
assemblies could be computed.
Note that the elastic (reversible) movement
$\Delta\boldsymbol{\xi}^{\text{(r)}}$ depends upon
the contact radius $a$ and hence upon the indentation
$\zeta$ and normal force $f^{\text{n}}$.
%
%
\section{DEM modeling details and quasi-static verification}\label{sec:DEM}
\begin{figure}
  \centering
  \includegraphics[scale=0.20]{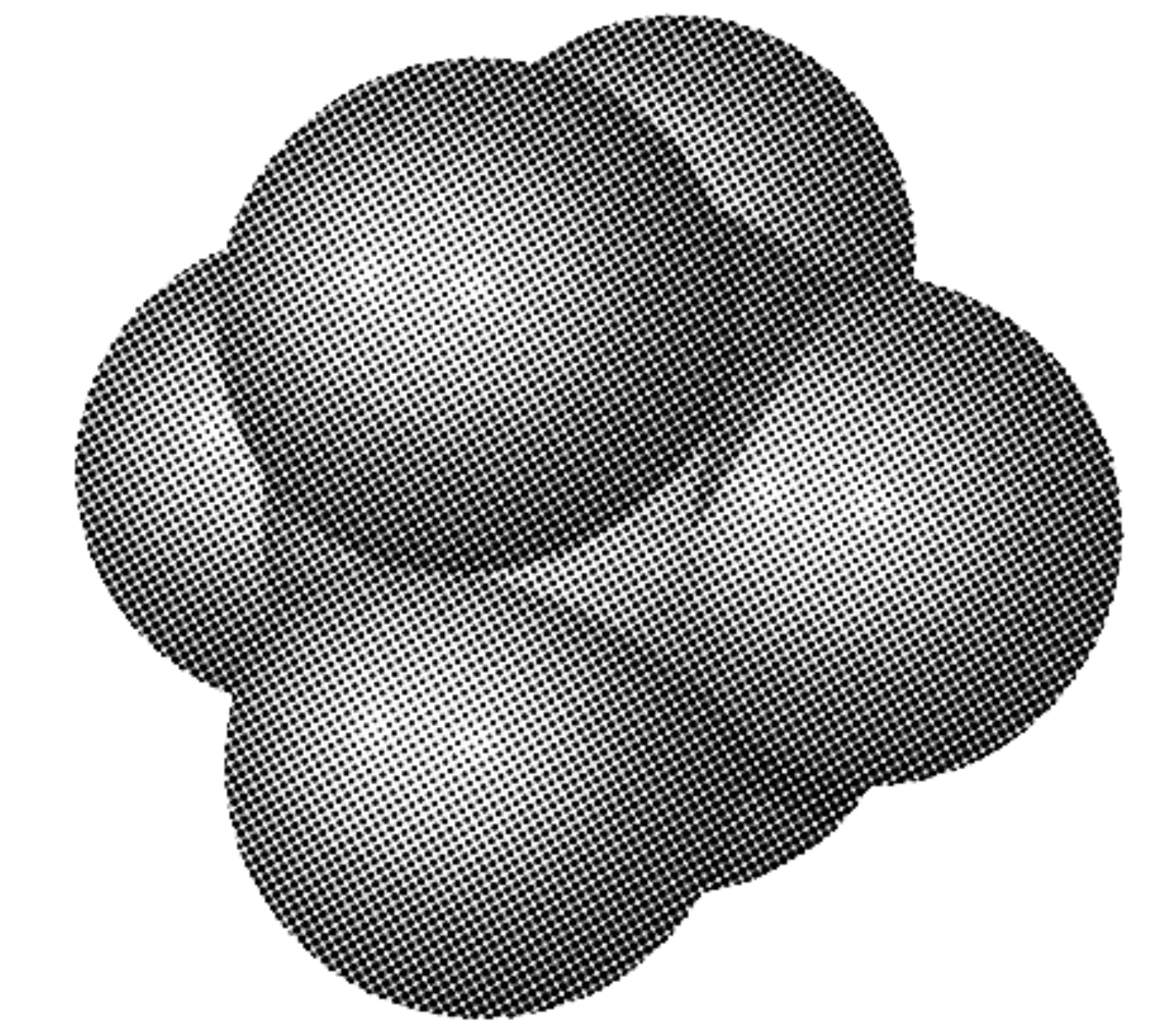}
  \caption{Sphere-cluster particle.
           \label{fig:spherecluster}}
\end{figure}

Discrete element
simulations were conducted with a cubical assembly of 10,648 particles contained
within periodic boundaries.
The simulations
were intended to produce a modest fidelity to the
bulk behavior of sands at low confining stress.
Because sphere assemblies produce unrealistic rolling between
particles and have a low bulk strength, we used a bumpy,
non-convex cluster shape for the particles:
a large central sphere with six smaller embedded outer
spheres in an octahedral arrangement (Fig.~\ref{fig:spherecluster}).
The use of non-spherical particles circumvents the need
of artificial measures to restrain the particle rotation
(for example, the use of rotational contact springs or
the direct restraint of particle rotations,
as in \cite{Calvetti:2003a,Plassiard:2009a}).
Through trial and error, we chose the radii of the central and outer
spheres so that the bulk behavior approximated that of Nevada
Sand, a standard poorly graded sand (SP) use in laboratory
and centrifuge testing programs
\cite{Arulmoli1992a,Kuhn:2014c}.
The particle size range was 0.074--0.28~mm,
with $D_{50}=0.165$~mm.
\par
To create assemblies with a range of densities,
we began with the particles sparsely and randomly
arranged within a cubic periodic cell.
With an initial low inter-particle friction coefficient
($\mu = 0.20$), the assembly was isotropically reduced in
dimension until it ``seized'' when a sufficiently complete
contact network had formed.
A series of 25 progressively denser assemblies were created
by repeatedly assigning random velocities to the particles
of the previous assembly and then further reducing the cell's
dimensions until it, too, had seized.
The 25 assemblies had void ratios ranging from
0.781 to 0.586 (solid fractions of 0.561 to 0.664).
The single assembly used in the paper
had void ratio 0.690
(solid fraction 0.592)
and approximates the behavior of Nevada Sand at a relative
density of 40\%.
After compaction, the assembly was allowed to quiesce with
friction coefficient $\mu=0.40$, which was then raised
to $\mu=0.55$ for the subsequent loading simulations.
\par
The particles are durable (non-breaking) and interact
only at their contacts.
Two contact models were used in the simulations: linear springs
with friction, and a Hertz, Cattaneo--Mindlin model,
as described in Section~\ref{sec:models} and~\ref{app:contacts}.
The Cattaneo--Mindlin model
was a full implementation of a
Hertz--Mindlin
contact between elastic-frictional bodies.
We used the
J\"{a}ger algorithm, which can model arbitrary sequences
of normal and tangential contact movements \cite{Kuhn:2011a}.
With the Cattaneo--Mindlin contacts,
the loading simulations were conducted with
an inter-particle friction coefficient
$\mu=0.55$,
particle shear modulus $G=29$~GPa, and Poisson ratio $\nu=0.15$.
\par
As has been noted by Alonso-Marroqu\'{i}n et al. \cite{AlonsoMarroquin:2005b}
and Froiio and Roux \cite{Froiio:2010a},
DEM simulations necessarily involve a compromise between scientific
intent and computational expedience.
Without proper care, DEM simulations can yield results
that are sensitive to strain rate, due to the particles' inertias
and to the
damping that is employed to stabilize
the particles' motions \cite{Suzuki:2014a}.
Our intent was to model behavior in which these effects
were minimized so that rate-independent behavior was attained.
To this end, we used a slow strain rate
(strain increments of $1\times 10^{-8}$)
and minimal viscous damping (2\% of critical damping).
\par
Several performance parameters were used to verify
the quasi-static and rate-independent nature of these probes.
The inertial number $I=\dot{\varepsilon}\sqrt{m/(pd)}$,
a relative measure of loading and inertial rates,
was about $1\times 10^{-11}$,
signifying nearly quasi-static loading
\cite{daCruz:2005a}.
During the incremental probes,
the average imbalance of force on a particle was less than
0.003\% of the average contact force
(parameter $\chi$ in \cite{Ng:2006a,Suzuki:2014a}).
The average kinetic energy of the particles was less than
$3\times 10^{-7}$ times the
elastic energy in the contacts.
With the very slow strain rates of the simulations,
doubling the strain increment from $1\times10^{-8}$
to $2\times10^{-8}$ had minimal effect on the
monotonic stress-strain response. 
Boundary movements were regulated so that any six of the
stress or strain components (or any six linear combination of these
components) could be controlled at desired rates.
When a stress component was controlled, it would typically remain
on target to within 0.001~Pa,
compared with the mean stress of 100~kPa and stress-probes
that produced stress changes $|d\sigma_{ij}|$ on the order
of 100~Pa.
As further verification of strain rate indifference during
loading,
we also conducted brief creep and stress relaxation tests in
which either the stress or the assembly boundaries were frozen
at the end of a stress probe.
Froiio and Roux \cite{Froiio:2010a} have noted a tendency
for an assembly to exhibit creep during small stress probes,
an inclination that can obscure the probe results.
During our creep tests, the strain rate was
0.5--3$\times 10^{-10}$, far less than the loading rate of $1\times10^{-8}$.
During stress relaxation with zero strain rate, the stress changed at
a rate of less than 4\% of that typically measured during
the incremental stress probes that were used in this
study.
The strain increment of $2\times 10^{-6}$ and strain steps of
$1\times 10^{-8}$ entailed 200 steps.
Although small fluctuations were noted in the first few steps,
the advance of stress and strain were fairly uniform within
the 200 strain steps.
Moreover,
the envelopes in Figs.~\ref{fig:piplane1},
\ref{fig:piplane2}, and~\ref{fig:Rendulic} are quite smooth,
even near the origin of these plots,
where the strain parts are minuscule, and at high magnification,
indicating that random errors are negligible.
All of these measurements indicate that the behavior in the
simulations was nearly quasi-static and independent of loading rate.
\par
\pagebreak
%

\end{document}